\documentclass[11pt,a4paper]{article}
\usepackage{jheppub}
\usepackage{amsmath,amssymb,amsfonts}
\usepackage{bm}
\usepackage{comment}
\usepackage{booktabs}

\usepackage{orcidlink}

\numberwithin{equation}{section}

\newcommand{\nd}{{\rm d}}
\newcommand{\ep}{\epsilon}
\newcommand{\tr}{{\rm T}}
\newcommand{\nn}{\\}

\title{Integration-By-Parts and \texorpdfstring{\textsf{d}\,\textsf{log}}{dlog} differential equations for strongly mixed cosmological correlators}
\author[c]{Pratik Chattopadhyay\,\orcidlink{0000-0002-5131-6334}}
\author[a,b]{Jiaqi Chen\,\orcidlink{0000-0003-3537-3846}}  
\author[d]{and Ye Zhou\,\orcidlink{0009-0004-7050-7736}}

\affiliation[a]{Beijing Key Laboratory of Optical Detection Technology for Oil and Gas, China University of Petroleum-Beijing, Beijing 102249, China}
\affiliation[b]{Basic Research Center for Energy Interdisciplinary, College of Science, China University of Petroleum-Beijing, Beijing 102249, China}
\affiliation[c]{Department of Physics, Ariel University, Ariel 40700, Israel}
\affiliation[d]{Institute of Systems and Physical Biology, Shenzhen Bay Laboratory, Shenzhen 518055, China}
\emailAdd{pratikpc@gmail.com}
\emailAdd{jiaqichen@cup.edu.cn}
\emailAdd{ye.zhou.horizon@gmail.com}

\abstract{
We establish universal formulas for recursive integration-by-parts reduction and $\mathrm{d}\log$-form differential equations to compute inflationary correlators in strongly mixed regimes, including cases with unequal sound speeds. In this regime, the solutions of the equations of motion are no longer Hankel functions. We evaluate all master integrals of the three-point correlators family and obtain the late-time bispectra induced by the four cubic interactions $\sigma^3$, $\dot{\pi}\sigma^2$, $\dot{\pi}^2\sigma$, and $\dot{\pi}^3$ for various relative sound speeds.
}


\begin{document}
\maketitle

\section{Introduction}
\label{sec:intro}
Inflation provides a natural setting in which quantum fluctuations are stretched to
cosmological scales and imprinted on primordial curvature perturbations. While the
power spectrum characterizes their nearly Gaussian component, higher-point
correlation functions, especially the bispectrum, probe the interactions and
particle content of the inflationary epoch
\cite{Maldacena:2002vr, Chen:2006nt}. Additional massive degrees of freedom can
leave distinctive non-Gaussian signatures, as in quasi-single-field inflation
\cite{Chen:2009we, Chen:2009zp, Noumi:2012vr}. In particular,
nonanalytic momentum dependence in squeezed configurations can carry information
about the masses and spins of particles coupled to the adiabatic fluctuation,
providing the central motivation for cosmological-collider physics
\cite{Arkani-Hamed:2015bza, Chen:2018xck}. Effective descriptions of single- and
multifield inflation furnish a systematic setting for such interactions
\cite{Cheung:2007st, Garcia-Saenz:2019njm, Pinol:2024arz}.
We consider a light adiabatic mode $\pi$ and a massive entropy mode $\sigma$,
with a derivative coupling already at quadratic order. Their late-time
correlators are computed using the in-in (Schwinger--Keldysh) formalism, which
expresses observables in terms of time integrals of mode functions and
propagators \cite{Weinberg:2005vy, Chen:2017ryl}. At weak mixing, the transfer
between the two fields can be organized perturbatively through mixed
propagators \cite{Chen:2017ryl}. At strong mixing, however, the quadratic
coupling must be resummed: the relevant building blocks are solutions of the
coupled mode equations and the associated dressed propagators. Obtaining the
full momentum dependence of their higher-point correlators therefore requires
methods that apply beyond special kinematic limits.

For tree-level cosmological correlators in the absence of strong quadratic
mixing, several complementary approaches have been developed. These include
bootstrap constructions \cite{Arkani-Hamed:2018kmz}, Mellin--Barnes
representations and Mellin-space bootstrap methods
\cite{Sleight:2019mgd, Sleight:2019hfp}, generalized IBP/DE methods
\cite{Chen:2023iix, Chen:2024glu} and their automation
\cite{Chen:2026iwf}, differential equations
and complete solutions for general trees \cite{Liu:2024str},
first-order differential equations for two-site massive correlators
\cite{Gasparotto:2024bku}, and spectral or gluing
constructions \cite{Werth:2024mjg, Grafe:2026avi}.

At loop level, massive-field cosmological correlators have been investigated
using direct numerical Schwinger--Keldysh integration
\cite{Wang:2021qez}, spectral decomposition and
bootstrap methods \cite{Xianyu:2022jwk}, and partial
Mellin--Barnes representations \cite{Qin:2024gtr}. IBP/DE methods have
also been developed for de Sitter loop-integral families
\cite{Chen:2026dqp, Chen:2026qmx}. Automated IBP-seed
generation for de Sitter loop diagrams is implemented in \textsc{dSIBP}
\cite{Chen:2026vll}.

For conformally coupled fields, simplified mode functions have enabled
relative-twisted-cohomology and intersection-theory methods for IBP reduction
and differential equations \cite{De:2023xue}, as well as
combinatorial kinematic-flow equations for arbitrary tree graphs
\cite{Arkani-Hamed:2023bsv, Arkani-Hamed:2023kig}.

For strongly mixed multifield systems, available analytic methods include
hypergeometric mode-function constructions
\cite{Huenupi:2026abj, Huenupi:2026aqc}, Schwinger-parameter representations
\cite{Pinol:2026xnl, Pinol:2026hey}, Mellin--Barnes decompositions
\cite{Wang:2026lff}, and Laplace/Heun representations \cite{Huenupi:2026jcv}.
Numerical cosmological flow provides a complementary approach
\cite{Werth:2023pfl, Pinol:2023oux, Werth:2024aui}, applicable to
both weakly and strongly mixed theories.

The generalized IBP/DE framework for independent Hankel modes motivates our
construction \cite{Chen:2023iix, Chen:2024glu}. This parallels flat-space IBP reduction to
master integrals \cite{Chetyrkin:1981qh, Maierhofer:2017gsa, Smirnov:2019qkx, Klappert:2020nbg} and differential-equation
methods \cite{Kotikov:1990kg, Kotikov:1991pm, Gehrmann:1999as, Henn:2013pwa}.
More broadly, recursive IBP relations and differential equations have
recently been constructed directly from the Regge--Wheeler--Zerilli equations
for the nonlinear response of Schwarzschild black holes at arbitrary
perturbative order \cite{Chen:2026wlo}. This demonstrates that,
in that setting, IBP relations can be obtained from the underlying equations
of motion without explicit analytic solutions for the mode functions. The
same principle motivates our treatment of strongly mixed inflationary modes,
whose individual field modes are not independent Hankel functions. The
existing Hankel-based reduction therefore cannot be applied directly. The
central task is to identify a closed first-order system for the dressed modes
and use it to construct an appropriate integral family.

Building on this principle, we develop a generalized IBP/DE framework
directly for the strongly mixed two-field system. Starting from the exact
coupled quadratic dynamics and dressed Schwinger--Keldysh propagators, we
construct a function basis obeying a first-order system with at most a
simple pole in conformal time. This structure yields closed generalized IBP
relations, an iterative reduction to master integrals, and
$\mathrm{d}\log$-form differential equations in momentum space for integral
families with arbitrary numbers of dressed legs. We fix the boundary data
using an auxiliary exponential factor and the asymptotic expansion of the
coupled modes, and then transport the solutions to physical momentum
configurations. Alternatively, we determine the boundary data directly at a fixed momentum configuration using Frobenius subtraction to extract the finite parts and residues of the master integrals. This provides an independent boundary construction for the subsequent momentum-space differential-equation transport, without introducing an auxiliary energy parameter. Finally, as an application and validation of the framework, we compute the
late-time bispectra induced by the four cubic interactions $\sigma^3$,
$\dot{\pi}\sigma^2$, $\dot{\pi}^2\sigma$, and $\dot{\pi}^3$ for isosceles and
general momentum triangles. To the best of our knowledge, the general-triangle
bispectra induced by these four interactions at unequal sound speeds have not
been computed before.

The paper is organized as follows. In section \ref{sec:feynman} we derive the strongly
mixed Feynman rules and define the integral family. In section \ref{sec:special}
we construct the simple-pole function system. Section \ref{sec:ibp} develops the
IBP reduction, and section \ref{sec:de} derives the $\nd\log$-form differential
equations. In section \ref{sec:boundary} we determine the boundary data and numerical
evaluation chain. Section \ref{sec:results} contains the bispectrum results, and
section \ref{sec:validation} presents the validation tests. We conclude in
section \ref{sec:summary}; appendix \ref{app:k0} collects auxiliary-family details.

\section{Strongly mixed propagation and Feynman rules}
\label{sec:feynman}

In this section we derive the Feynman rules for the strongly mixed theory and identify the resulting time-integral family, following the dressed Schwinger-Keldysh construction of \cite{Huenupi:2026aqc}.

\subsection{The coupled quadratic theory}
\label{sec:quad}

We consider an adiabatic fluctuation $\pi$ coupled to a scalar entropy field $\sigma$ in de Sitter space, with metric $\nd s^2=a_{\rm sc}^2(\tau)(-\nd\tau^2+\nd\bm x^2)$ and scale factor $a_{\rm sc}(\tau)=-1/(H\tau)$. We treat the quadratic mixing exactly and the cubic interactions to first order. The quadratic action is
\begin{align}
S_2=\int \nd t_c\,\nd^3x\,a_{\rm sc}^3\left[\frac{\dot\pi^2+\dot\sigma^2}{2}-\frac{c_\pi^2(\nabla\pi)^2+c_\sigma^2(\nabla\sigma)^2}{2a_{\rm sc}^2}-\frac{H^2\mu^2\sigma^2}{2}-H\lambda\,\dot\pi\,\sigma\right],
\label{eq:S2}
\end{align}
Here $t_c$ is cosmic time, $\mu$ is the entropy mass in units of $H$, and $\lambda$ is the dimensionless mixing strength. All quadratic parameters are constant, and we work in the regime
\begin{align}
c_\pi,c_\sigma>0,\qquad \lambda\neq 0,\qquad S\equiv \mu^2+\lambda^2>\frac94,\qquad \gamma=\sqrt{S-\frac94},\qquad \Delta=\frac32+i\gamma .
\label{eq:params}
\end{align}
The mixing can be absorbed by completing the square,
\begin{align}
\frac{\dot\pi^2}{2}-H\lambda\,\dot\pi\sigma-\frac{H^2\mu^2\sigma^2}{2}=\frac{(\dot\pi-H\lambda\sigma)^2}{2}-\frac{H^2S\sigma^2}{2},
\label{eq:sq}
\end{align}
so that the effective entropy mass squared in this form of the action is $H^2S$.

Writing a canonically normalized Fourier mode as $Hk^{-3/2}(u(z),v(z))$ with $z=-k\tau$ and $D=z\partial_z$, the coupled equations of motion following from \eqref{eq:S2} are
\begin{align}
\left[D(D-3)+c_\pi^2 z^2\right]u&=\lambda(3-D)v,\nn\\
\left[D(D-3)+c_\sigma^2 z^2+\mu^2\right]v&=\lambda D u .
\label{eq:coupled}
\end{align}
It is useful to introduce the canonical combinations
\begin{align}
Y=(Y_1,Y_2,Y_3,Y_4)^{\tr},\qquad
Y_1=u,\quad Y_2=v,\quad Y_3=Du+\lambda v,\quad Y_4=Dv,\qquad
Y_5=-Du .
\label{eq:Y}
\end{align}
Here $Y_5$ is the mode function of the normalized velocity $\dot\pi/H$. Since the momentum conjugate to $\pi$ is $\Pi_\pi=a_{\rm sc}^3(\dot\pi-H\lambda\sigma)$, the mode function of $\Pi_\pi/(a_{\rm sc}^3H)$ is simply $-Y_3$.
The two second-order equations \eqref{eq:coupled} are equivalent to the first-order system
\begin{align}
DY=(L_0+z^2L_2)Y,
\label{eq:firstorder}
\end{align}
with constant matrices
\begin{align}
L_0=\begin{pmatrix}0&-\lambda&1&0\\0&0&0&1\\0&0&3&0\\0&-S&\lambda&3\end{pmatrix},\qquad
L_2=\begin{pmatrix}0&0&0&0\\0&0&0&0\\-c_\pi^2&0&0&0\\0&-c_\sigma^2&0&0\end{pmatrix}.
\label{eq:L0L2}
\end{align}

\subsection{Dressed Schwinger-Keldysh propagators}
\label{sec:dressed}

Let $\mathbb K_0$ denote the unmixed Schwinger-Keldysh kernel and $\mathbb G_0=\mathbb K_0^{-1}$ its propagator. Writing the full quadratic kernel as $\mathbb K=\mathbb K_0-\mathcal M$, where $\mathcal M$ represents the mixing insertion, the dressed propagator $\mathbb G=\mathbb K^{-1}$ admits the formal expansion
\begin{align}
\mathbb G=\mathbb G_0+\mathbb G_0*\mathcal M*\mathbb G_0+\mathbb G_0*\mathcal M*\mathbb G_0*\mathcal M*\mathbb G_0+\cdots,
\label{eq:resum}
\end{align}
where $*$ denotes integration over intermediate times and summation over field and contour indices. Since each insertion changes the field species, $G_{\pi\pi}$ contains an even number and $G_{\pi\sigma}$ an odd number of insertions. The series \eqref{eq:resum} is a formal diagrammatic expansion of the exact inverse; we use the exact coupled modes below and do not expand in $\lambda$.

Let $Y_\alpha$, $\alpha=1,2$, be the two positive-frequency Bunch-Davies (BD) modes with unit canonical normalization. With $\phi_\pi=\pi$, $\phi_\sigma=\sigma$, $z=-k\tau$ and $z'=-k\tau'$, the exact Wightman matrix of the original fields is
\begin{align}
\mathbf G^>(k;\tau,\tau')=\left[\langle\phi_{A,\mathbf k}(\tau)\phi_{B,-\mathbf k}(\tau')\rangle'\right]_{A,B=\pi,\sigma}=\frac{H^2}{k^3}\sum_{\alpha=1}^2\begin{pmatrix}u_\alpha(z)\overline{u_\alpha(z')}&u_\alpha(z)\overline{v_\alpha(z')}\\ v_\alpha(z)\overline{u_\alpha(z')}&v_\alpha(z)\overline{v_\alpha(z')}\end{pmatrix},
\label{eq:wightman}
\end{align}
where the prime removes $(2\pi)^3\delta^3(\mathbf k+\mathbf k')$. The oppositely ordered function is $G^<_{AB}(k;\tau,\tau')=G^>_{BA}(k;\tau',\tau)$. The four Schwinger-Keldysh bulk-to-bulk propagators, with $+$ and $-$ labelling the forward and backward contour branches, are
\begin{align}
\mathbf G_{++}&=\theta(\tau-\tau')\mathbf G^>+\theta(\tau'-\tau)\mathbf G^<,\qquad \mathbf G_{+-}=\mathbf G^<,\nn\\
\mathbf G_{--}&=\theta(\tau'-\tau)\mathbf G^>+\theta(\tau-\tau')\mathbf G^<,\qquad \mathbf G_{-+}=\mathbf G^>.
\label{eq:skprop}
\end{align}
These propagators contain the quadratic mixing to all orders, so no separate transfer vertices are required. We consider the cubic interactions
\begin{align}
S_3=-\int\nd\tau\,\nd^3x\,a_{\rm sc}^4\sum_{r=0}^{3}\frac{g_r}{r!(3-r)!}\left(\frac{\dot\pi}{H}\right)^r\sigma^{3-r}.
\label{eq:S3}
\end{align}
For a three-point function with external $\pi$ fields, the factorials cancel the Wick multiplicity for each of the $\binom{3}{r}$ assignments of the velocity insertions to the external legs. At first order in the cubic couplings, each assignment carries a vertex factor $-ig_r a_{\rm sc}^4\nd\tau$ on the $+$ branch and $+ig_r a_{\rm sc}^4\nd\tau$ on the $-$ branch, with each factor of $\dot\pi/H$ acting as $(a_{\rm sc}H)^{-1}\partial_\tau$ on the corresponding propagator. We evaluate the four contributions separately at unit coupling.

For a cubic contact vertex with late-time $\pi$ external fields, only the late-time $\pi$ row of the propagator matrix and its vertex-time derivatives are needed. The response and the power are
\begin{align}
Y_{\rm resp}(z)=\sum_{\alpha=1}^2 u_\alpha(0)\overline{Y_\alpha(z)},\qquad P=\sum_{\alpha=1}^2|u_\alpha(0)|^2 .
\label{eq:response}
\end{align}
The late-time two-point function is then $\langle\pi_{\mathbf k}(0)\pi_{-\mathbf k}(0)\rangle'=H^2P/k^3$. From here on, $Y_i$ denote the response components, while the individual BD modes carry the index $\alpha$. The bulk-to-boundary propagators on the $+$ branch are then
\begin{align}
\mathcal K_\pi(k,\tau)=\frac{H^2}{k^3}u(z),\quad \mathcal K_\sigma(k,\tau)=\frac{H^2}{k^3}v(z),\quad \mathcal K_h(k,\tau)=\frac{1}{a_{\rm sc}H}\partial_\tau\mathcal K_\pi(k,\tau)=-\frac{H^2}{k^3}Du(z),
\label{eq:bulkboundary}
\end{align}
and the $-$ branch uses the complex-conjugate external lines at real times.

\subsection{The general integral family}
\label{sec:intfamily}

A tree-level contact diagram involves a single time integral over products of the response functions and their derivatives. Using \eqref{eq:bulkboundary}, the integrand for the cubic interactions \eqref{eq:S3} is built from $v$ and $Y_5$, evaluated at the respective leg momenta, together with powers of the scale factor. More generally, polynomial insertions of the fields and their first time derivatives can be expressed in terms of the four components of $Y$.

This is the strong-mixing analogue of the de Sitter integral family introduced in \cite{Chen:2023iix}, where the integrand contains Hankel functions, their derivatives and a separate exponential factor $e^{ik_0\tau}$. In our case the oscillations are encoded in the coupled mode functions themselves. An auxiliary exponential factor will be introduced in section~\ref{sec:auxexp} to construct boundary data. The first step towards an IBP reduction is to find a basis of one-leg functions whose derivatives relate integrals at adjacent time weights.

\section{The special functions and their simple-pole equation}
\label{sec:special}

The key to extending the Hankel-function IBP construction of \cite{Chen:2023iix} to strongly mixed propagation is to find a basis in which the differential equation consists of a constant matrix and a simple pole.


\subsection{Constructing a simple-pole basis}
\label{sec:simplepole}
Written in terms of ordinary derivatives, \eqref{eq:firstorder} reads $\nd Y/\nd z=(L_0/z+zL_2)Y$. The term $zL_2$ couples integrals whose time weights differ by two. To eliminate it, we rescale the field components by $z^{-1}$ and the momentum components by $z^{-2}$, allowing shifts proportional to $v$ in the latter. Taking $W_3=[Du+(\lambda-\eta)v]/z^2$ and $W_4=(Dv-\xi v)/z^2$, the cancellation of the resulting $1/z^2$ terms requires $\eta(3-\xi)=0$ and $\lambda\eta+\xi(3-\xi)=S$. Since $\xi(3-\xi)\leq 9/4<S$ for real $\xi$, the branch $\eta=0$ is excluded, leaving $\xi=3$ and $\eta=S/\lambda$. This leads to the following rational change from the original mode vector $Y$ in \eqref{eq:Y} to the deformed function basis $W$:
\begin{align}
W=\begin{pmatrix}
z^{-1}&0&0&0\\[6pt]
0&z^{-1}&0&0\\[6pt]
0&-\dfrac{S}{\lambda z^2}&z^{-2}&0\\[6pt]
0&-\dfrac3{z^2}&0&z^{-2}
\end{pmatrix}
\begin{pmatrix}Y_1\\Y_2\\Y_3\\Y_4\end{pmatrix}
=\begin{pmatrix}\dfrac{u}{z}\\[6pt] \dfrac{v}{z}\\[6pt] \dfrac{Du+(\lambda-S/\lambda)v}{z^2}\\[6pt] \dfrac{Dv-3v}{z^2}\end{pmatrix}.
\label{eq:Wdef}
\end{align}
In this basis, the first-order system \eqref{eq:firstorder} becomes
\begin{align}
\frac{\nd W}{\nd z}=\left(A+\frac{B}{z}\right)W,
\label{eq:Weq}
\end{align}
with the constant matrices
\begin{align}
A=\begin{pmatrix}0&0&1&0\\0&0&0&1\\-c_\pi^2&0&0&0\\0&-c_\sigma^2&0&0\end{pmatrix},\qquad
B=\begin{pmatrix}-1&\dfrac{\mu^2}{\lambda}&0&0\\[6pt]0&2&0&0\\[6pt]0&0&1&-\dfrac{S}{\lambda}\\[6pt]0&0&\lambda&-2\end{pmatrix}.
\label{eq:AB}
\end{align}

The four components $W_a$, $a\in\{1,2,3,4\}$, form the one-leg function basis for the strongly mixed theory, generalizing the two Hankel kernels of \cite{Chen:2023iix}. The field and velocity insertions entering the cubic vertices are recovered from \eqref{eq:Wdef} as
\begin{align}
u=zW_1,\qquad v=zW_2,\qquad Y_5=-\frac{\mu^2}{\lambda}zW_2-z^2W_3 .
\label{eq:inverse}
\end{align}
The transformation \eqref{eq:Wdef} is defined for $\lambda\neq0$. At $\lambda=0$, the mode equations \eqref{eq:coupled} decouple, and the Hankel-function basis of \cite{Chen:2023iix} can be used directly.

\subsection{Why the simple pole matters}
\label{sec:whypole}

The matrices $A$ and $B$ govern the asymptotic behaviour of the special functions at large and small $z$, respectively. The eigenvalues $\pm ic_\pi$ and $\pm ic_\sigma$ of $A$ determine the leading exponential factors at large $z$, while the simple pole at $z=0$ gives a regular-singular Frobenius structure. The four branches of the original system \eqref{eq:firstorder} near the origin are
\begin{align}
Y_\beta(z)=z^\beta\sum_{j\geq0}y_{\beta,j}z^{2j},\qquad \beta\in\{0,3,\Delta,\bar\Delta\},
\label{eq:frobenius}
\end{align}
where $\bar\Delta=3/2-i\gamma$. The transformation \eqref{eq:Wdef} gives $W=O(z^{-1})$ as $z\to0$. At large $z$, the Bunch--Davies response carries the exponential factors $e^{-ic_\pi z}$ and $e^{-ic_\sigma z}$. We examine the individual Frobenius branches in section~\ref{sec:modeasym}.

For the IBP reduction, I need the derivatives of the special functions with respect to both time and momentum. Using $z=-k\tau$ directly in \eqref{eq:Weq} gives
\begin{align}
\partial_\tau W_a(-k\tau)&=\sum_{c=1}^4\left(-kA_{ac}+\frac{B_{ac}}{\tau}\right)W_c(-k\tau),\nonumber\\
\partial_k W_a(-k\tau)&=\sum_{c=1}^4\left(-\tau A_{ac}+\frac{B_{ac}}{k}\right)W_c(-k\tau).
\label{eq:Wdiff}
\end{align}
Both derivatives close on the four components of $W$. In the total-derivative IBP identity, the constant term $A$ raises the integral weight from $n$ to $n+1$, while the pole term $B/z$ preserves it. The same two weights appear under momentum differentiation. A $1/z^2$ term would introduce the additional weight $n-1$, whereas the linear term of the original system introduces $n+2$. The transformed system therefore leads directly to the two-weight recursion derived in the next section.

\section{Generalized IBP relations and iterative reduction}
\label{sec:ibp}

We now construct the integral family generated by the special functions of section~\ref{sec:special}. Its IBP relations connect adjacent time weights and lead to an iterative reduction to master integrals, which we formulate for an arbitrary number of legs.

\subsection{The single-vertex integral family}
\label{sec:vtxfamily}

For $N$ legs, we use the physical momenta $k_i$ and conformal time $\tau$ directly:
\begin{align}
z_i=-k_i\tau .
\label{eq:momentummap}
\end{align}
The Bunch--Davies contour $\mathcal C$ runs from $-\infty$ to $0$ with the positive imaginary prescription; it can be rotated to $\tau=is$, $s>0$, where the response decays. For $N$ legs meeting at a single interaction vertex, we define the integral family
\begin{align}
\mathcal I^{(n)}_{\bm a}(\ep;\bm k)=\int_{\mathcal C}\nd\tau\,(-\tau)^{\ep+n-1}\prod_{i=1}^N W_{a_i}(-k_i\tau),\qquad a_i\in\{1,2,3,4\},\quad n\in\mathbb Z .
\label{eq:Ifamily}
\end{align}
Here $\bm a=(a_1,\ldots,a_N)$ specifies the component on each leg, while $n$ labels the integer shift in the time weight. We write $w=\ep+n$, so that the common convergence domain is ${\rm Re}\,w>N$: near the origin each $W_a$ is at most of order $\tau^{-1}$, while the rotated contour ensures convergence at infinity.

The cubic vertices select integer values of the time weight. From the first row of \eqref{eq:AB}, the derivative of the adiabatic response $u=zW_1$ is
\begin{align}
\frac{\nd u}{\nd z}=z\,W_3+\frac{\mu^2}{\lambda}\,W_2 .
\label{eq:upee}
\end{align}
With $z_i=-k_i\tau$, the field insertions are $v(z_i)=z_iW_2(z_i)$ and $Y_5(z_i)=-z_i[(\mu^2/\lambda)W_2(z_i)+z_iW_3(z_i)]$. For a vertex with $r$ velocity insertions, the Wick contractions sum over the $\binom{3}{r}$ choices of velocity legs. After factoring out the common factor $(-1)^r\prod_i z_i$, the remaining kernel can be written as
\begin{align}
\mathcal V_r(\tau;\bm k)
&=\sum_{\substack{H\subset\{1,2,3\}\\|H|=r}}
\prod_{i\in H}\left[\frac{\mu^2}{\lambda}W_2(z_i)+z_iW_3(z_i)\right]
\prod_{j\notin H}W_2(z_j)\nonumber\\
&=\sum_{m=0}^{r}\binom{3-m}{r-m}\left(\frac{\mu^2}{\lambda}\right)^{r-m}
\sum_{\substack{M\subset\{1,2,3\}\\|M|=m}}
\prod_{i\in M}z_iW_3(z_i)
\prod_{j\notin M}W_2(z_j),
\qquad r=0,1,2,3 .
\label{eq:velocity}
\end{align}
Here $H$ specifies the velocity legs, while $M$ specifies the legs carrying $W_3$. For a fixed choice of $M$, the remaining $r-m$ velocity insertions can be placed on any of the other $3-m$ legs, giving the multiplicity $\binom{3-m}{r-m}$. The factor $\prod_i z_i$ combines with the vertex measure $a_{\rm sc}^4\nd\tau\propto\tau^{-4}\nd\tau$ to give the basic measure $\tau^{-1}\nd\tau$, while each $W_3$ insertion supplies one additional power of $\tau$. Thus a term containing $m$ factors of $W_3$ belongs to weight $n=m$. In particular, the four interactions $\sigma^3$, $\dot\pi\sigma^2$, $\dot\pi^2\sigma$ and $\dot\pi^3$ involve weights $0$, $0,1$, $0,1,2$ and $0,1,2,3$, respectively. The $r=3$ case gives the eight-term velocity-vertex combination. The family is initially defined in the common convergence domain, and the physical vertices are obtained by analytic continuation to $\ep=0$, with the regulator kept finite until all terms of each vertex have been assembled.

\subsection{IBP relations}
\label{sec:ibprel}

Within the common convergence domain, the boundary terms vanish when we integrate the total derivative of $(-\tau)^{\ep+n}\prod_i W_{a_i}(-k_i\tau)$. Using the first of \eqref{eq:Wdiff}, we obtain
\begin{align}
0=(\ep+n)\,\mathcal I^{(n)}_{\bm a}+\sum_{i=1}^N\sum_{c=1}^4 B_{a_i c}\,\mathcal I^{(n)}_{\bm a[i\to c]}+\sum_{i=1}^N\sum_{c=1}^4k_i A_{a_i c}\,\mathcal I^{(n+1)}_{\bm a[i\to c]},
\label{eq:IBP}
\end{align}
Here $\bm a[i\to c]$ denotes the multi-index obtained by replacing $a_i$ with $c$. The relation involves only the four functions $W_a$ and connects integrals at adjacent time weights, generalizing the Hankel-function IBP relation of \cite{Chen:2023iix}.

Differentiating with respect to a leg scale $k_i$ at fixed $\ep$, $n$, and the other leg scales, and using the second of \eqref{eq:Wdiff}, gives
\begin{align}
\partial_{k_i}\mathcal I^{(n)}_{\bm a}=\frac{1}{k_i}\sum_{c=1}^4 B_{a_i c}\,\mathcal I^{(n)}_{\bm a[i\to c]}+\sum_{c=1}^4 A_{a_i c}\,\mathcal I^{(n+1)}_{\bm a[i\to c]} .
\label{eq:momdiff}
\end{align}
The term proportional to $A$ raises the weight by one, while the $B$ term preserves it.

\subsection{Iterative reduction}
\label{sec:iterative}

Equation \eqref{eq:IBP} connects only adjacent time weights, allowing the integral family to be reduced recursively. To make this structure explicit, we collect the $4^N$ integrals at each weight into a vector $\bm I^{(n)}$, ordered by the multi-index $\bm a$.

Let $A^{(i)}$ and $B^{(i)}$ act as the explicit one-leg matrices $A$ and $B$ in \eqref{eq:AB} on the $i$th tensor factor and as the identity elsewhere:
\begin{align}
A^{(i)}&=\mathbb I_4^{\otimes(i-1)}\otimes A\otimes\mathbb I_4^{\otimes(N-i)},\qquad
B^{(i)}=\mathbb I_4^{\otimes(i-1)}\otimes B\otimes\mathbb I_4^{\otimes(N-i)} .
\label{eq:slotoperators}
\end{align}
With the physical leg momenta $k_i$ in \eqref{eq:momentummap}, we define
\begin{align}
A_\Sigma(\bm k)=\sum_{i=1}^Nk_i A^{(i)},\qquad B_\Sigma=\sum_{i=1}^N B^{(i)} .
\label{eq:AsumBsum}
\end{align}
Then \eqref{eq:IBP} reads
\begin{align}
\left[(\ep+n)\mathbb I_{4^N}+B_\Sigma\right]\bm I^{(n)}+A_\Sigma(\bm k)\,\bm I^{(n+1)}=0 .
\label{eq:IBPmatrix}
\end{align}
The downward and upward shifts are
\begin{align}
\bm I^{(n-1)}=A_-(n)\,\bm I^{(n)},\qquad \bm I^{(n+1)}=A_+(n)\,\bm I^{(n)},
\label{eq:shift}
\end{align}
with
\begin{align}
A_-(n)=-\left[(\ep+n-1)\mathbb I_{4^N}+B_\Sigma\right]^{-1}A_\Sigma(\bm k),\qquad
A_+(n)=-A_\Sigma(\bm k)^{-1}\left[(\ep+n)\mathbb I_{4^N}+B_\Sigma\right].
\label{eq:Apm}
\end{align}
The regulator must be retained in the weight matrices throughout the reduction. The downward shift from weight $n$ to $n-1$ requires $(\ep+n-1)\mathbb I_{4^N}+B_\Sigma$ to be invertible, while the upward shift requires $A_\Sigma(\bm k)$ to be invertible. Since the matrices $A^{(i)}$ commute, the eigenvalues of $A_\Sigma$ are written directly as
\begin{align}
\operatorname{spec}A_\Sigma
=\left\{i\sum_{j=1}^N\omega_jk_j:\;
\omega_j\in\{\pm c_\pi,\pm c_\sigma\}\right\}.
\label{eq:frequencyspectrum}
\end{align}
The four choices for each leg are
\begin{align}
(\omega_i,f_i)\in\left\{(c_\pi,\pi),\,(-c_\pi,\pi),\,
(c_\sigma,\sigma),\,(-c_\sigma,\sigma)\right\}.
\label{eq:frequencychoices}
\end{align}
Each $\omega_i$ is a signed frequency, and its field tag $f_i$ is retained even when the sound speeds coincide. The upward shift is available away from the frequency hypersurfaces $\sum_j\omega_jk_j=0$. Away from these hypersurfaces and the resonant weights, the two shifts express the entire integral family in terms of the $4^N$ reference integrals $\bm I^{(0)}$, which we take as our master integrals.

The inverses in \eqref{eq:Apm} are obtained from one-leg data, without a generic inversion of a $4^N\times4^N$ matrix. Explicitly, the one-leg matrices are diagonalized as
\begin{align}
A&=U_A\operatorname{diag}(ic_\pi,-ic_\pi,ic_\sigma,-ic_\sigma)U_A^{-1},\nonumber\\
B&=U_B\operatorname{diag}\left(-1,2,-\tfrac12+i\gamma,-\tfrac12-i\gamma\right)U_B^{-1},
\label{eq:onelegdiagonalization}
\end{align}
For $N$ legs, write $U_{A,N}=U_A^{\otimes N}$ and $U_{B,N}=U_B^{\otimes N}$. The explicit one-leg diagonalizers are
\begin{align}
U_A=\begin{pmatrix}
1&1&0&0\\
0&0&1&1\\
ic_\pi&-ic_\pi&0&0\\
0&0&ic_\sigma&-ic_\sigma
\end{pmatrix},\qquad
U_B=\begin{pmatrix}
1&\mu^2&0&0\\
0&3\lambda&0&0\\
0&0&S/\lambda&S/\lambda\\
0&0&\tfrac32-i\gamma&\tfrac32+i\gamma
\end{pmatrix}.
\label{eq:onelegdiagonalizers}
\end{align}
The diagonalized $N$-leg matrices are
\begin{align}
\widetilde A_\Sigma&=U_{A,N}^{-1}A_\Sigma U_{A,N}
=\operatorname{diag}_{\bm\omega}\left(i\sum_{j=1}^N\omega_jk_j\right),\nonumber\\
\widetilde B_\Sigma&=U_{B,N}^{-1}B_\Sigma U_{B,N}
=\operatorname{diag}_{\bm\rho}\left(\sum_{j=1}^N\rho_j\right),
\label{eq:diagonalizedoperators}
\end{align}
where $\rho_j\in\{-1,2,-\tfrac12\pm i\gamma\}$. The required inverses are then
\begin{align}
A_\Sigma^{-1}&=U_{A,N}\widetilde A_\Sigma^{-1}U_{A,N}^{-1},\nonumber\\
\left(c\mathbb I_{4^N}+B_\Sigma\right)^{-1}
&=U_{B,N}\left(c\mathbb I_{4^N}+\widetilde B_\Sigma\right)^{-1}U_{B,N}^{-1}.
\label{eq:inverseoperators}
\end{align}
For the downward shift in \eqref{eq:Apm}, $c=\ep+n-1$. Thus all inverse operations reduce to diagonalizing the explicit $4\times4$ matrices \eqref{eq:onelegdiagonalizers} and forming tensor products; the zeros of $\widetilde A_\Sigma$ and $c\mathbb I_{4^N}+\widetilde B_\Sigma$ give the frequency hypersurfaces and resonant weights, respectively.

Explicit three-leg examples of these weight resonances are collected in section~\ref{sec:singleindicial}.

\section{\texorpdfstring{$\nd\log$}{dlog}-form differential equations}
\label{sec:de}

The IBP relations allow us to express the momentum derivatives entirely in terms of integrals at a fixed weight. The spectrum of the frequency matrix then gives a $\nd\log$ representation of the differential equations, extending the construction of \cite{Chen:2023iix} to strongly mixed propagation.

\subsection{Momentum-space differential equations}
\label{sec:dederive}

At a fixed weight $w=\ep+n$, define the weight matrix
\begin{align}
M_w=w\mathbb I_{4^N}+B_\Sigma .
\label{eq:weightmatrix}
\end{align}
The momentum derivative \eqref{eq:momdiff} and the upward IBP relation \eqref{eq:Apm} give
\begin{align}
\partial_{k_i}\bm I^{(n)}=\frac{B^{(i)}}{k_i}\bm I^{(n)}+A^{(i)}\bm I^{(n+1)},\qquad
\bm I^{(n+1)}=-A_\Sigma^{-1}M_w\bm I^{(n)}.
\label{eq:de1}
\end{align}
Away from the zeros of $\det A_\Sigma$, this yields the closed system
\begin{align}
\partial_{k_i}\bm I^{(n)}=\Omega_{k_i}(w)\bm I^{(n)},\qquad
\Omega_{k_i}(w)=\frac{B^{(i)}}{k_i}-A^{(i)}A_\Sigma^{-1}M_w .
\label{eq:dematrix}
\end{align}
The master-integral equations correspond to $n=0$, or $w=\ep$.

\subsection{\texorpdfstring{$\nd\log$}{dlog} structure}
\label{sec:dlog}

To write \eqref{eq:dematrix} in $\nd\log$ form, I need $A_\Sigma^{-1}$ and its differential. Both are immediate in the basis diagonalizing $A_\Sigma$. Using the four tagged frequency choices in \eqref{eq:frequencychoices}, the slot matrices are simultaneously diagonal, with entries
\begin{align}
A^{(i)}P_{\bm\omega}=i\omega_iP_{\bm\omega},\qquad
A_\Sigma P_{\bm\omega}=i\left(\sum_{j=1}^N\omega_jk_j\right)P_{\bm\omega}.
\label{eq:frequencyprojectors}
\end{align}
Here $\bm\omega=(\omega_1,\ldots,\omega_N)$ labels a joint frequency eigenspace, and $P_{\bm\omega}$ denotes the spectral projector onto it.
Away from the zero-frequency hypersurfaces, I use the inverse in \eqref{eq:inverseoperators}. On each joint eigenspace, its differential gives
\begin{align}
\sum_i A^{(i)}A_\Sigma^{-1}\nd k_i
=\sum_{\bm\omega}P_{\bm\omega}\,
\nd\log\left(\sum_{j=1}^N\omega_jk_j\right).
\label{eq:inversefrequencydifferential}
\end{align}
Substituting into \eqref{eq:dematrix}, we obtain
\begin{align}
\nd\bm I^{(n)}=\Omega(w)\bm I^{(n)},\qquad
\Omega(w)=\sum_i B^{(i)}\nd\log k_i
-\sum_{\bm\omega}P_{\bm\omega}M_w\,
\nd\log\left(\sum_{j=1}^N\omega_jk_j\right).
\label{eq:dlog}
\end{align}
The alphabet consists of the soft letters $k_i$ and the linear forms $\sum_j\omega_jk_j$. We use $\ell$ for any one of these forms when discussing a generic frequency hypersurface. The residues are independent of the momenta and affine in $w$. Proportional frequency letters share the same logarithmic differential, so their projectors are summed in the corresponding residue. For three legs, the frequency letters are
\begin{align}
\ell_{\bm\omega}=\omega_1k_1+\omega_2k_2+\omega_3k_3 .
\label{eq:threelegletters}
\end{align}
Up to overall signs, the complete three-leg alphabet, including the soft letters, is
\begin{align}
\mathcal A_3=\{k_1,k_2,k_3\}\cup
\bigcup_{f_1,f_2,f_3\in\{\pi,\sigma\}}
\left\{\begin{gathered}
c_{f_1}k_1+c_{f_2}k_2+c_{f_3}k_3,\\
c_{f_1}k_1+c_{f_2}k_2-c_{f_3}k_3,\\
c_{f_1}k_1-c_{f_2}k_2+c_{f_3}k_3,\\
c_{f_1}k_1-c_{f_2}k_2-c_{f_3}k_3
\end{gathered}\right\}.
\label{eq:threelegalphabet}
\end{align}
The field-dependent coefficients $c_{f_i}$ generate additional frequency hypersurfaces at unequal sound speeds.

\subsection{Block decomposition}
\label{sec:blockdecomposition}

The block structures of $A$ and $B$ in \eqref{eq:AB} induce a grading of the IBP system. Let
\begin{align}
Q=\operatorname{diag}(1,1,-1,-1),\qquad Q_N=Q^{\otimes N}.
\end{align}
Since $QAQ=-A$ and $QBQ=B$, conjugation by $Q_N$ reverses the sign of $A_\Sigma$ and leaves $M_w$ invariant. The IBP relation \eqref{eq:IBPmatrix} therefore separates into the two eigensectors of the weight-dependent parity $(-1)^nQ_N$:
\begin{align}
[(-1)^nQ_N,M_w]=0,\qquad
(-1)^{n+1}Q_NA_\Sigma=A_\Sigma(-1)^nQ_N .
\label{eq:blockdecomposition}
\end{align}
Both sectors contain $4^N/2$ components, giving two $32$-component systems at three legs.

\section{Asymptotics and indicial structure}
\label{sec:modeasym}

The asymptotic behaviour of the special functions controls the convergence of the time integrals and provides the expansion coefficients needed for their boundary data. We first examine the two endpoints of the one-leg equation and then determine the indicial spectra of the momentum-space connection \eqref{eq:dlog}.

\subsection{The single-leg indicial equation}
\label{sec:singleindicial}

The point $z=0$ is a regular singular point of \eqref{eq:Weq}. Substituting the Frobenius expansion $W=z^\rho\sum_{j\geq0}w_jz^j$ gives
\begin{align}
Bw_0=\rho w_0,\qquad
[(\rho+j)\mathbb I_4-B]w_j=Aw_{j-1},\qquad j\geq1 .
\label{eq:Wrecursion}
\end{align}
The two diagonal blocks of $B$ give the indicial polynomial
\begin{align}
\det(\rho\mathbb I_4-B)=(\rho+1)(\rho-2)(\rho^2+\rho+S-2).
\end{align}
The four exponents are therefore
\begin{align}
\mathcal E=\left\{-1,\ 2,\ -\tfrac12+i\gamma,\ -\tfrac12-i\gamma\right\}.
\label{eq:Wexp}
\end{align}
For three legs, sums of these single-leg exponents include
\begin{align}
0&=(-1)+(-1)+2,\qquad
-2=\left(-\tfrac12-i\gamma\right)+\left(-\tfrac12+i\gamma\right)+(-1),\nonumber\\
-3&=(-1)+(-1)+(-1).
\label{eq:resonance}
\end{align}
These combinations occur for every $\gamma$, making the downward steps to $n=0,2,3$ resonant at $\ep=0$. We retain the regulator throughout the reduction and take the limit $\ep\to0$ after assembling each physical vertex, whose finiteness follows from section~\ref{sec:modeendpoint}. These weight resonances are distinct from the frequency hypersurfaces $\sum_j\omega_jk_j=0$.

The two integer exponents differ by three, but the corresponding branches of the original system \eqref{eq:firstorder} have expansions in even powers of $z$, as in \eqref{eq:frobenius}. The resonance thus causes no obstruction, and all four branches admit Frobenius expansions without logarithms.

\subsection{Endpoint behaviour of the mode functions}
\label{sec:modeendpoint}

The transformation \eqref{eq:Wdef} relates the four original-field branches $\beta=0,3,\Delta,\bar\Delta$ to the special functions. On the $\beta=0$ branch, the leading coefficient of $v$ vanishes, while on both integer branches the leading terms in $Du+(\lambda-S/\lambda)v$ and $Dv-3v$ cancel. The resulting powers are collected in table~\ref{tab:Wbranchpowers}.

\begin{table}[htbp]
\centering
\small
\setlength{\tabcolsep}{6pt}
\begin{tabular}{lcccc}
\toprule
Original-field branch & $W_1$ & $W_2$ & $W_3$ & $W_4$ \\
\midrule
$\beta=0$ & $z^{-1}$ & $z$ & $1$ & $1$ \\
$\beta=3$ & $z^2$ & $z^2$ & $z^3$ & $z^3$ \\
$\beta=\Delta$ & $z^{\Delta-1}$ & $z^{\Delta-1}$ & $z^{\Delta-2}$ & $z^{\Delta-2}$ \\
$\beta=\bar\Delta$ & $z^{\bar\Delta-1}$ & $z^{\bar\Delta-1}$ & $z^{\bar\Delta-2}$ & $z^{\bar\Delta-2}$ \\
\bottomrule
\end{tabular}
\caption{Generic leading powers of the special functions on the four Frobenius branches near $z=0$, with $\Delta=\tfrac32+i\gamma$ and $\bar\Delta=\tfrac32-i\gamma$.}
\label{tab:Wbranchpowers}
\end{table}

Each entry is followed by corrections in even powers of $z$. Along a fixed ray, the complete response therefore satisfies $W=O(z^{-1})$, with $W_2=O(z^{1/2})$ and $W_3,W_4=O(z^{-1/2})$. The physical insertions \eqref{eq:inverse} then obey
\begin{align}
u=O(1),\qquad v,Y_5=O(z^{3/2}).
\label{eq:physicalendpoint}
\end{align}

At large $z$, we use the frequency eigenbasis of $A$ labelled by \eqref{eq:frequencychoices}.
In this basis, all diagonal entries of $B$ vanish. For unequal sound speeds, the asymptotic expansions consequently have leading exponential factors $e^{\pm i c_\pi z}$ and $e^{\pm i c_\sigma z}$, with inverse-power corrections. At equal sound speeds $c_\pi=c_\sigma=c$, the two frequency eigenvectors of each sign span a degenerate subspace. In a suitably normalized frequency basis, the corresponding block of $B$ can be written as
\begin{align}
C=\frac{\lambda}{2}
\begin{pmatrix}
0&-1\\
1&0
\end{pmatrix}.
\label{eq:degeneratefrequencyblock}
\end{align}
Its eigenvalues $\pm i\lambda/2$ give the additional powers $z^{\pm i\lambda/2}$ multiplying $e^{\pm i c z}$. In both cases, the Bunch--Davies response selects the negative exponential frequencies and decays along the Euclidean contour.

\subsection{Indicial structure of the integral family}
\label{sec:nspectrum}

These spectra identify where weight reduction or momentum transport becomes singular. Since $B_\Sigma$ in \eqref{eq:AsumBsum} is a sum of matrices acting on separate tensor factors, its spectrum is obtained by adding the single-leg exponents,
\begin{align}
\operatorname{spec}(B_\Sigma)=
\left\{\sum_{i=1}^N\rho_i:\rho_i\in\mathcal E\right\}.
\label{eq:weightspectrum}
\end{align}
These sums determine the singular weights of the downward IBP relation \eqref{eq:Apm}, including the three-leg resonances at $n=0,2,3$ when $\ep=0$.

The momentum-space indicial spectra follow from the residues of \eqref{eq:dlog}. For $N\geq2$, at a generic point of the soft hypersurface $k_i=0$, the residue is $B^{(i)}$, with spectrum $\mathcal E$. At a generic point of a frequency hypersurface $\ell=0$, away from intersections with other singular hypersurfaces, we combine the projectors of all proportional letters,
\begin{align}
P_\ell=\sum_{\bm\omega:\,\sum_j\omega_jk_j\propto\ell}P_{\bm\omega},
\qquad R_\ell=-P_\ell M_w .
\label{eq:frequencyresidue}
\end{align}
Let $d_\ell=\operatorname{rank}P_\ell$. In the decomposition into the image and kernel of $P_\ell$, the residue takes the block-triangular form
\begin{align}
R_\ell=
\begin{pmatrix}
-w\mathbb I_{d_\ell}-C_\ell & -P_\ell B_\Sigma(\mathbb I-P_\ell)\\
0&0
\end{pmatrix},
\qquad
C_\ell=\left.P_\ell B_\Sigma P_\ell\right|_{\operatorname{im}P_\ell}.
\end{align}
Its characteristic polynomial is therefore
\begin{align}
\det(\nu\mathbb I_{4^N}-R_\ell)
=\nu^{4^N-d_\ell}\det\left[(\nu+w)\mathbb I_{d_\ell}+C_\ell\right].
\label{eq:nleg-spectrum}
\end{align}

For unequal sound speeds, two distinct frequency states associated with proportional letters must differ in at least two tensor slots. Since each term of $B_\Sigma$ acts on only one slot, the off-diagonal entries of $C_\ell$ vanish in the tensor-product frequency basis. Its diagonal entries vanish by the one-leg result above. Thus $C_\ell=0$, and the residue has exponents $-w$ with multiplicity $d_\ell$ and zero with multiplicity $4^N-d_\ell$.

At equal sound speeds, each frequency sign pattern contains $2^N$ field assignments, and each hypersurface groups a sign pattern with its opposite. Within either pattern, $C_\ell$ is the tensor sum of $N$ copies of the block \eqref{eq:degeneratefrequencyblock}. The corresponding exponents are
\begin{align}
\nu_j=-w+\frac{i\lambda}{2}(N-2j),\qquad j=0,\ldots,N,
\label{eq:equalspeedexponents}
\end{align}
with multiplicities $2\binom Nj$, together with $4^N-2^{N+1}$ zero exponents. In particular, at three legs the frequency projector on a folding wall has rank four at unequal sound speeds and rank sixteen at equal sound speeds. For real noninteger $w$, the residues at generic points of these frequency hypersurfaces are diagonalizable and their transverse Frobenius expansions contain no logarithms.

At intersections of singular hypersurfaces, the local residue along a given path is the sum of the individual residues weighted by the orders of vanishing of their letters. Consider, for example, the isosceles soft limit $\bm k=(k,k,q)$ as $q\to0$ at fixed $k$. Using the projectors in \eqref{eq:frequencyprojectors}, define
\begin{align}
P_{\rm h}=\sum_{\substack{\bm\omega=(\omega_1,\omega_2)\\\omega_1+\omega_2=0}}P_{\bm\omega},
\qquad
B_{\rm h}=B\otimes\mathbb I_4+\mathbb I_4\otimes B .
\label{eq:hardprojector}
\end{align}
Thus $P_{\rm h}$ projects onto the zero-frequency states of the two hard legs. The combined residue is
\begin{align}
R_{q=0}
&=B^{(3)}-(P_{\rm h}\otimes\mathbb I_4)M_w\nonumber\\
&=(\mathbb I_{16}-P_{\rm h})\otimes B
-P_{\rm h}(w\mathbb I_{16}+B_{\rm h})\otimes\mathbb I_4 .
\label{eq:isoscelessoftresidue}
\end{align}
In the decomposition of the hard-leg space into the image and kernel of $P_{\rm h}$, this residue is block triangular. Its diagonal blocks contribute the single-leg spectrum $\mathcal E$ and the spectrum of the compressed hard-leg weight matrix, respectively. The compression of $B_{\rm h}$ vanishes at unequal sound speeds and has eigenvalues $0,\pm i\lambda$ at equal sound speeds. The sets of distinct soft exponents are therefore
\begin{align}
\operatorname{spec}(R_{q=0})=
\begin{cases}
\mathcal E\cup\{-w\}, & c_\pi\neq c_\sigma,\\
\mathcal E\cup\{-w,-w+i\lambda,-w-i\lambda\}, & c_\pi=c_\sigma .
\end{cases}
\label{eq:softspectrum}
\end{align}

The residue spectra determine the allowed local exponents of the connection, while the boundary conditions select the combination of local solutions entering the physical integral. The Bunch--Davies integrals themselves have a simpler analytic structure in the positive-momentum domain. Their exponential decay on the Euclidean contour persists uniformly in a sufficiently narrow sector about that ray, so the integrals \eqref{eq:Ifamily} are analytic near every strictly positive momentum configuration in the common convergence domain. At the physical cubic weights, each term in \eqref{eq:velocity} is also integrable at the origin: the estimates \eqref{eq:physicalendpoint} give an integrand of order $|\tau|^{1/2}$ after including the vertex measure. The cubic bispectra are consequently regular across the frequency hypersurfaces within the positive-momentum domain, including the folding walls. These hypersurfaces are singularities of the differential connection, but not of the Bunch--Davies contact integrals.

\section{Boundary conditions}
\label{sec:boundary}
\label{sec:chain}

The differential equations determine the integral family once its boundary data are fixed. In the unmixed construction of \cite{Chen:2024glu}, these data are obtained at large auxiliary energy, where the time integrals reduce to Gamma-function series. The strongly mixed family \eqref{eq:Ifamily} has no separate exponential energy factor, but introducing one allows us to recover the same construction. An alternative is to evaluate the integrals directly at a momentum anchor. Both approaches start from the normalized Bunch--Davies response.

\subsection{Bunch--Davies normalization}
\label{sec:bd}

The positive-frequency Bunch--Davies modes are selected by their decay along the Euclidean ray $z=is$, $s>0$, where \eqref{eq:Weq} becomes
\begin{align}
\frac{\nd W}{\nd s}=\left(iA+\frac{B}{s}\right)W .
\label{eq:euclideanmode}
\end{align}
The two decaying solutions span the positive-frequency subspace. On the positive real axis, the canonical inner product is
\begin{align}
\langle Y_\alpha,Y_\beta\rangle_{\rm can}=-iz^{-3}Y_\alpha^\dagger\mathsf JY_\beta,\qquad
\mathsf J=\begin{pmatrix}0&\mathbb I_2\\-\mathbb I_2&0\end{pmatrix}.
\label{eq:symplectic}
\end{align}
The identities $L_0^{\tr}\mathsf J+\mathsf JL_0=3\mathsf J$ and $L_2^{\tr}\mathsf J+\mathsf JL_2=0$ ensure its conservation. The corresponding bilinear form $z^{-3}Y_\alpha^{\tr}\mathsf JY_\beta$ vanishes on the decaying subspace.

Let $\mathsf U(z)=(\widetilde Y_\alpha(z))_{\alpha=1,2}$ be a basis of the positive-frequency subspace. Its Gram matrix and late-time adiabatic amplitudes are
\begin{align}
G=-iz^{-3}\mathsf U^\dagger(z)\mathsf J\mathsf U(z),\qquad
\bm u_0=\big(\widetilde u_1(0),\widetilde u_2(0)\big)^{\tr}.
\label{eq:KGgram}
\end{align}
The matrix $G$ is positive definite and independent of $z$. The response \eqref{eq:response} can then be written directly in this basis as
\begin{align}
Y_{\rm resp}(z)=\overline{\mathsf U(\bar z)}\,G^{-\tr}\bm u_0 .
\label{eq:gramresponse}
\end{align}
To see this, write an orthonormal basis as $\mathsf U T$, with $T^\dagger GT=\mathbb I_2$. Its late-time amplitudes are $T^{\tr}\bm u_0$, and inserting them into \eqref{eq:response} gives \eqref{eq:gramresponse} through $\overline T T^{\tr}=G^{-\tr}$. The response is therefore independent of the choice of basis and, unlike the individual positive-frequency modes, decays along the ray $z=-is$.

Numerically, we approximate the decaying subspace at a large Euclidean starting point by the stable eigenspace of $iA+B/s$, integrate towards the origin and continue to the positive real axis. Matching the resulting modes to the four Frobenius branches and applying \eqref{eq:gramresponse} gives the response series
\begin{align}
W_a(z)=\sum_{\rho\in\mathcal E}\sum_{j=0}^{\infty}c_{a,\rho,j}z^{\rho+j},
\label{eq:responseseries}
\end{align}
with $\mathcal E$ given in \eqref{eq:Wexp}. The coefficients depend on $(\mu,\lambda,c_\pi,c_\sigma)$ and are shared by all momentum configurations at fixed model parameters.

\subsection{Boundary data from an auxiliary exponential}
\label{sec:auxexp}

For the numerical boundary construction, we restore dimensionless variables by
\begin{align}
t=-k_*\tau,\qquad \kappa_i=k_i/k_* .
\label{eq:numericalscaling}
\end{align}
The overall factor $k_*^{-w}$ generated by this rescaling is restored when converting back to the physical integral \eqref{eq:Ifamily}. We introduce an auxiliary energy $k_0$ and define
\begin{align}
J^{(n)}_{\bm a}(\ep;\bm\kappa;k_0)=\int_0^\infty\nd t\,t^{w-1}e^{ik_0t}\prod_{i=1}^N W_{a_i}(\kappa_i t),\qquad w=\ep+n .
\label{eq:Jfamily}
\end{align}
For ${\rm Re}\,w>N$ and ${\rm Im}\,k_0>0$, the integral converges on the real axis. Its limit at $k_0=0$ defines the numerical form of the Bunch--Davies integral:
\begin{align}
\mathcal I^{(n)}_{\bm a}(\ep;\bm\kappa)
&\equiv J^{(n)}_{\bm a}(\ep;\bm\kappa;0),\qquad
\mathcal I^{(n)}_{\bm a}(\ep;\bm k)
=k_*^{-w}\mathcal I^{(n)}_{\bm a}(\ep;\bm\kappa).
\label{eq:numericalintegral}
\end{align}
Accordingly, the numerical formulas below use $A_\Sigma(\bm\kappa)$, $\Omega_{\kappa_i}$, and the frequency projectors evaluated with $\bm k=\bm\kappa$.

The exponential factor shifts the frequency matrix in the IBP relation by $ik_0\mathbb I_{4^N}$, while differentiation with respect to $k_0$ raises the weight:
\begin{align}
M_w\bm J^{(n)}+(A_\Sigma+ik_0\mathbb I_{4^N})\bm J^{(n+1)}=0,
\label{eq:JIBP}\\
\partial_{k_0}\bm J^{(n)}=i\bm J^{(n+1)}.
\label{eq:Jk0}
\end{align}
Here $A_\Sigma$ and $M_w$ are given in \eqref{eq:AsumBsum} and \eqref{eq:weightmatrix}, respectively, and $\omega_j$ are the signed frequencies defined in \eqref{eq:frequencyspectrum}. Using the frequency projectors \eqref{eq:frequencyprojectors}, we obtain
\begin{align}
\partial_{k_0}\bm J^{(n)}
&=-i(A_\Sigma+ik_0\mathbb I_{4^N})^{-1}M_w\bm J^{(n)}\nonumber\\
&=-\sum_{\bm\omega}\frac{P_{\bm\omega}M_w}{k_0+\sum_{j=1}^N\omega_j\kappa_j}\bm J^{(n)}.
\label{eq:auxiliaryconnection}
\end{align}
The auxiliary connection thus has poles at the signed frequency thresholds $k_0=-\sum_j\omega_j\kappa_j$.

The boundary vector can be constructed at a positive real energy $K>c_{\max}\sum_i\kappa_i$, with $c_{\max}=\max(c_\pi,c_\sigma)$. Rotating the contour to $t=is$ gives the damping factor $e^{-Ks}$, and each power in the product of the Frobenius series \eqref{eq:responseseries} can be integrated using
\begin{align}
\int_0^{i\infty}\nd t\,e^{iKt}t^{\alpha-1}
=\frac{\Gamma(\alpha)}{(-iK)^\alpha},\qquad {\rm Re}\,\alpha>0,
\label{eq:gamma}
\end{align}
where the powers are continued from the positive real axis. We obtain
\begin{align}
J^{(n)}_{\bm a}(\ep;\bm\kappa;K)
=\sum_{\substack{\rho_1,\ldots,\rho_N\in\mathcal E\\j_1,\ldots,j_N\geq0}}
\frac{\Gamma\left(w+\sum_i(\rho_i+j_i)\right)}
{(-iK)^{w+\sum_i(\rho_i+j_i)}}
\prod_{i=1}^N c_{a_i,\rho_i,j_i}\kappa_i^{\rho_i+j_i}.
\label{eq:gammaboundary}
\end{align}
The convergence of this series follows from the recurrence \eqref{eq:Wrecursion}. In the frequency eigenbasis, the coefficients at order $j$ are bounded by $c_{\max}^j/j!$ times a polynomial in $j$. Grouping the terms in \eqref{eq:gammaboundary} by their total order $m=\sum_i j_i$ therefore gives a bound proportional to $(c_{\max}\sum_i\kappa_i/K)^m$ times a polynomial in $m$. The Gamma series converges absolutely for the stated range of $K$. The equality with \eqref{eq:Jfamily} can first be established for purely imaginary $k_0$ with sufficiently large positive imaginary part, where termwise integration on the real axis is absolutely convergent. Analytic continuation then gives \eqref{eq:gammaboundary} as the boundary value at real $K$ approached from the upper half plane.

Starting from this boundary, we integrate \eqref{eq:auxiliaryconnection} through the upper half of the complex $k_0$ plane to $k_0=0$. The path follows the analytic branch defined by the convergent real-axis integral \eqref{eq:Jfamily}. Near $k_0=0$, rotation to $t=-is$ gives an integral that is analytic in $k_0$, since the Bunch--Davies response supplies exponential decay. The physical endpoint is therefore regular for the defining integral, even at a zero of a frequency letter. Such zeros are singular points of the connection and require separate treatment in its numerical solution.

For the five-sound-speed calculation in section~\ref{sec:results}, the one-leg modes are initialized at $s=30$ and matched to the Frobenius basis at $z=1/2$. The one-leg series are constructed using $90$-digit arithmetic, retaining orders through $j=24$ and applying coefficient-pruning thresholds of $10^{-30}$. At each momentum configuration and regulator value, the Gamma series supplies a $64$-component boundary vector at $K=100$ and weight $w=4+\ep$. We transport this vector to $k_0=0$ at nonsingular configurations and apply the downward IBP relation \eqref{eq:Apm} to obtain the four sectors $n=3,2,1,0$. These sectors are shared by all four cubic vertices, whose physical contributions are obtained from \eqref{eq:velocity} in the limit $\ep\to0$. The numerical extraction is described in section~\ref{sec:errorbudget}.

\subsection{Direct boundary data at a momentum anchor}
\label{sec:anchor}

A second approach is to evaluate the boundary vector directly at a fixed momentum configuration where $A_\Sigma$ is invertible, and then transport it using \eqref{eq:dlog}. In the common convergence domain, the integrals can be evaluated along the Euclidean contour $t=-is$. Their continuation to other weights follows by subtracting the Frobenius expansion near the origin.

Let $f_{\bm a}(t)=\prod_iW_{a_i}(\kappa_i t)$ and let $\sum_p C_p t^p$ contain the terms of its small-$t$ expansion needed for convergence at the desired weight. Splitting the contour at $t_c=-is_c$, with $s_c>0$, gives
\begin{align}
\mathcal I^{(n)}_{\bm a}(\ep;\bm\kappa)
={}&\int_{t_c}^{-i\infty}\nd t\,t^{w-1}f_{\bm a}(t)
+\int_0^{t_c}\nd t\,t^{w-1}\left[f_{\bm a}(t)-\sum_p C_p t^p\right]\nonumber\\
&+\sum_p C_p\frac{t_c^{w+p}}{w+p}.
\label{eq:anchorfp}
\end{align}
The first two terms are analytic near the chosen weight, while the last term supplies the possible simple poles at $w=-p$. This determines both the residues and the finite parts of the integral vector from the same normalized response.

At an ordinary momentum configuration, the upward IBP relation \eqref{eq:Apm} connects the weight sectors without inverting $M_w$. If a sector has a pole at $\ep=0$, its residue contributes to the finite part of the next sector through the explicit $\ep$ dependence of the shift matrix. The corresponding recursions for the weight shifts and momentum transport are given in appendix~\ref{app:ladder}.

The direct and auxiliary constructions give the same integrals in their common convergence domain and hence the same meromorphic continuation in the weight. They provide two ways of fixing the boundary data for the momentum-space differential equations.
\section{Results}
\label{sec:results}

We apply the IBP--DE construction to the four cubic interactions in \eqref{eq:S3}, computing their late-time bispectra over general momentum triangles for five relative sound speeds.

\subsection{Numerical setup}
\label{sec:numericalsetup}

For the interaction with $r$ velocity insertions, we define the reduced bispectrum coefficient
\begin{align}
I_r(\ep;\bm\kappa)&=(-1)^r\left(\prod_{i=1}^3\kappa_i\right)\int_0^\infty\nd t\,t^{\ep-1}\mathcal V_r(t;\bm\kappa),\nonumber\\
B_r(\bm\kappa)&=2\,\operatorname{Im}I_r(0;\bm\kappa),
\label{eq:bispectrumdefinition}
\end{align}
where $\mathcal V_r$ is the vertex kernel in \eqref{eq:velocity}, evaluated at $z_i=\kappa_i t$, and the $t$ contour runs from $0$ to $+\infty$ with the lower-half-plane prescription. With $k_*=k_3$, the bulk-to-boundary propagators \eqref{eq:bulkboundary} and the two branches of the in-in contour give
\begin{align}
\left\langle\pi_{\mathbf k_1}\pi_{\mathbf k_2}\pi_{\mathbf k_3}\right\rangle'_r
=\frac{g_rH^2k_3^3}{(k_1k_2k_3)^3}B_r(\bm\kappa).
\label{eq:physicalbispectrum}
\end{align}
We present the dimensionless coefficients $B_r$ for the four channels separately. The corresponding momentum-weighted bispectrum is
\begin{align}
(k_1k_2k_3)^2\left\langle\pi_{\mathbf k_1}\pi_{\mathbf k_2}\pi_{\mathbf k_3}\right\rangle'_r
=g_rH^2\frac{B_r}{\kappa_1\kappa_2}.
\label{eq:weightedbispectrum}
\end{align}

We fix $(\mu,\lambda)=(3/2,5/2)$, so that $\gamma=5/2$, and set $c_\pi=1$. The relative sound speed is varied over
\begin{align}
r_c=\frac{c_\sigma}{c_\pi}\in\left\{\frac45,\frac9{10},1,\frac98,\frac54\right\}.
\label{eq:soundspeedscan}
\end{align}
This choice varies the propagation speeds while keeping the late-time scaling exponents fixed.

The permutation symmetry of the bispectrum allows us to restrict the momentum ratios to the ordered triangle
\begin{align}
0<\kappa_1\leq\kappa_2\leq\kappa_3=1,\qquad \kappa_1+\kappa_2\geq1 .
\label{eq:triangledomain}
\end{align}
The vertices $(0,1)$, $(1/2,1/2)$ and $(1,1)$ of its closure correspond to the squeezed, folded and equilateral configurations, respectively. We sample this domain on the grid
\begin{align}
\kappa_1=\frac12+\frac{i-j}{60},\qquad
\kappa_2=\frac12+\frac{i+j}{60},\qquad
i\geq1,\quad j\geq0,\quad i+j\leq30,
\label{eq:momentumgrid}
\end{align}
with integer $i,j$, giving $465$ nominal configurations per sound speed. The folded edge $i=0$ is excluded.

The zeros of the frequency letters are identified by exact rational arithmetic before integration. Across the five sound speeds, $20$ isolated grid points on frequency hypersurfaces are left unevaluated, while $12$ points on lines of constant $\kappa_2-\kappa_1$ are displaced by $1/60$ in that difference at fixed $\kappa_1+\kappa_2$. The displaced points are evaluated at their effective momentum coordinates. This gives $2305$ momentum configurations across the five models. At each configuration and regulator value, the auxiliary construction of section~\ref{sec:auxexp} supplies the integral vectors shared by all four vertices, yielding $9220$ reduced bispectrum values after regulator removal.

\subsection{Bispectrum surfaces and sound-speed dependence}
\label{sec:surface}

\begin{figure}[htbp]
\centering
\includegraphics[width=1.2\textwidth]{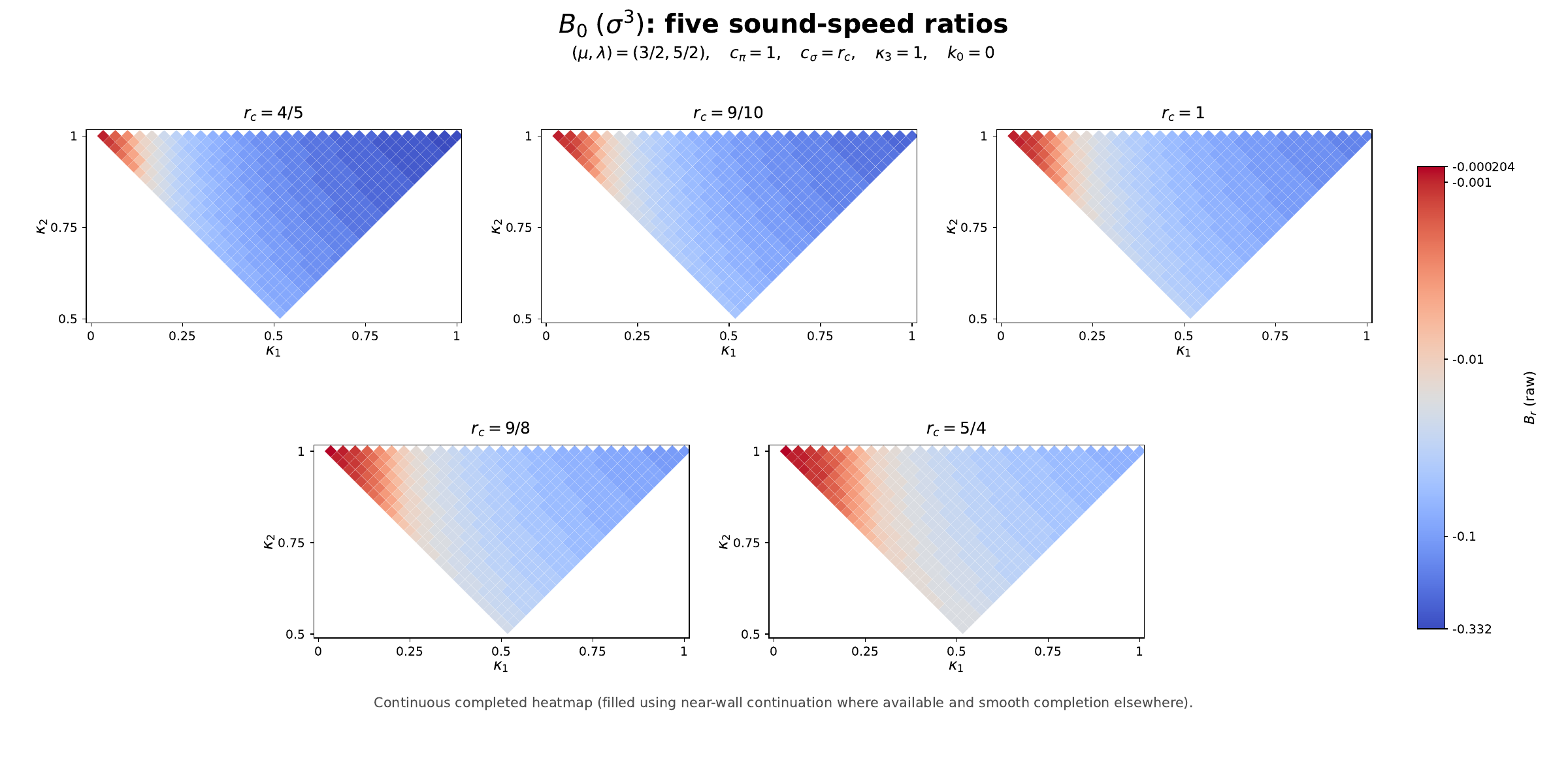}
\caption{Bispectrum $B_0$ generated by the cubic $\sigma^3$ interaction for five sound-speed ratios
$r_c=\{4/5,\,9/10,\,1,\,9/8,\,5/4\}$.
The bispectrum is shown over the momentum-triangle domain
parameterized by $(\kappa_1,\kappa_2)$, with $\kappa_3=1$,
for fixed parameters $(\mu,\lambda)=(3/2,5/2)$ and $c_\pi=1$.
A common nonlinear color scale is used across all five panels
to facilitate comparison of the absolute magnitudes.
Blue indicates more negative values of $B_0$, while red
indicates values closer to zero.}
\label{fig:bispectrum_rc45}
\end{figure}

\begin{figure}[htbp]
\centering
\includegraphics[width=1.2\textwidth]{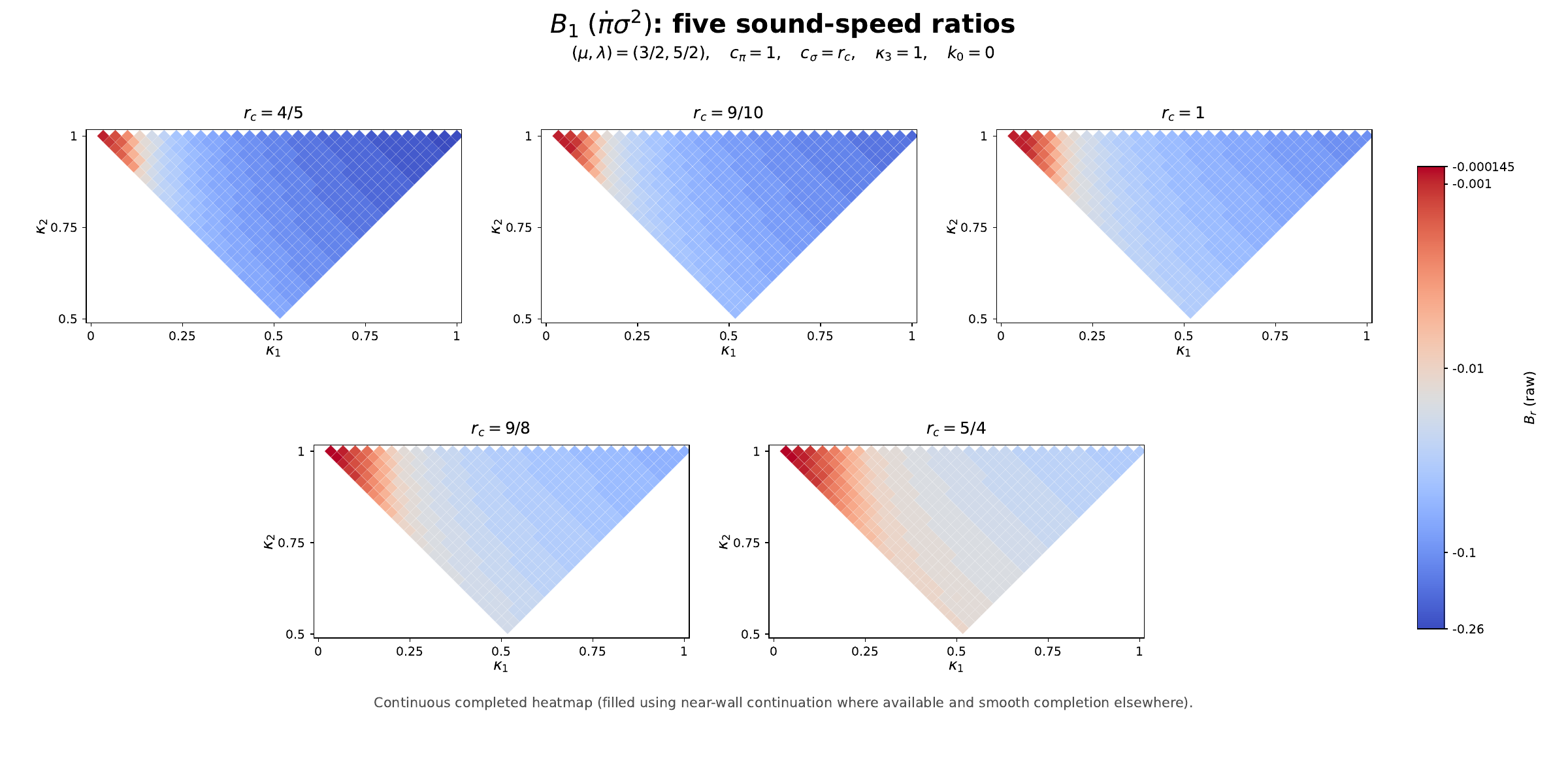}
\caption{Bispectrum $B_1$ arising from the $\dot{\pi}\sigma^2$
interaction for five sound-speed ratios
$r_c=\{4/5,\,9/10,\,1,\,9/8,\,5/4\}$. The model parameters, momentum coordinates and panel ordering are as in figure~\ref{fig:bispectrum_rc45}.}
\label{fig:bispectrum_rc910}
\end{figure}

\begin{figure}[htbp]
\centering
\includegraphics[width=1.2\textwidth]{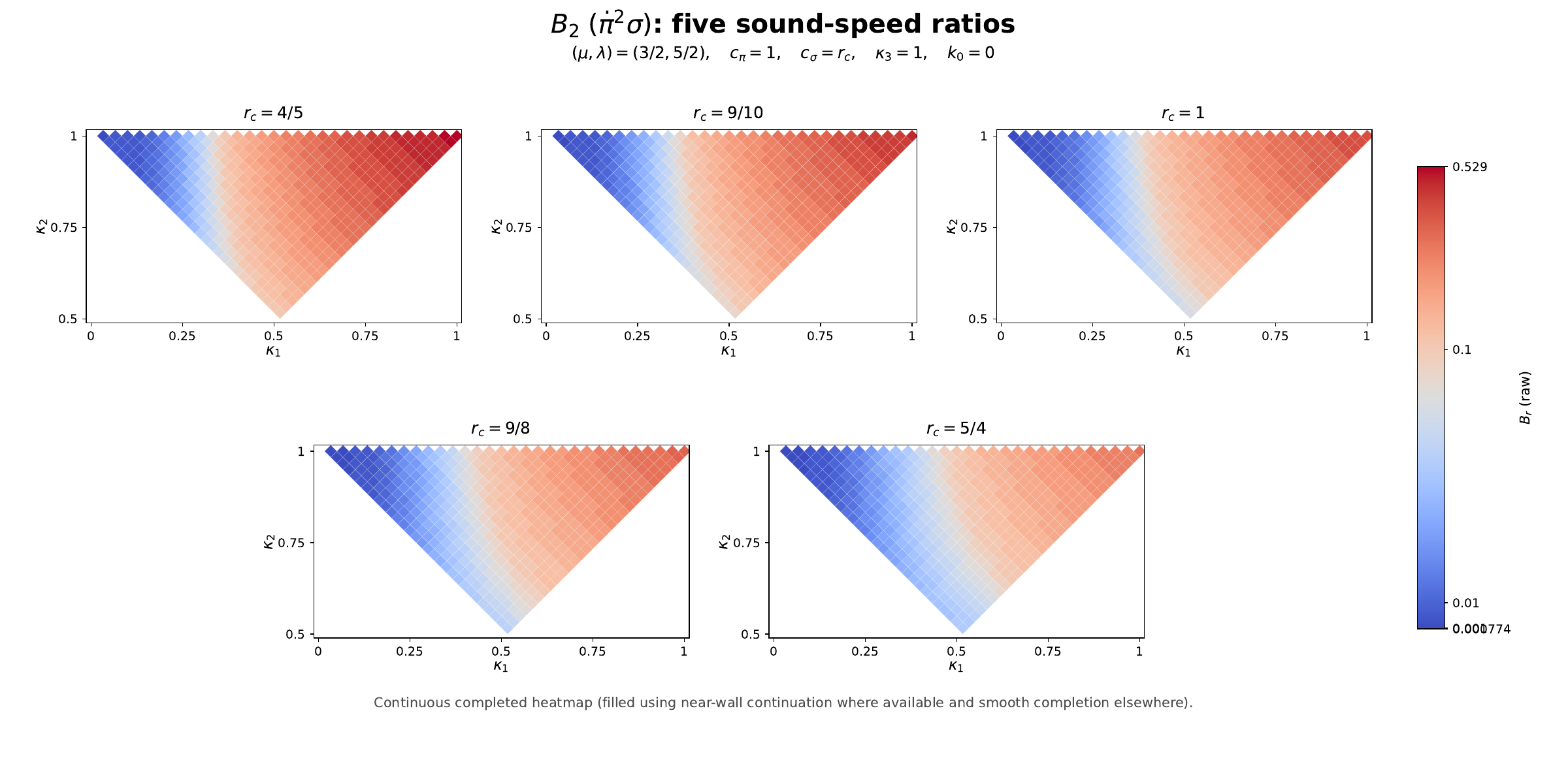}
\caption{Bispectrum $B_2$ arising from the $\dot{\pi}^2\sigma$
interaction for five sound-speed ratios
$r_c=\{4/5,\,9/10,\,1,\,9/8,\,5/4\}$. The remaining model parameters, momentum coordinates and panel ordering are as in figure~\ref{fig:bispectrum_rc45}.}
\label{fig:bispectrum_rc1}
\end{figure}

\begin{figure}[htbp]
\centering
\includegraphics[width=1.2\textwidth]{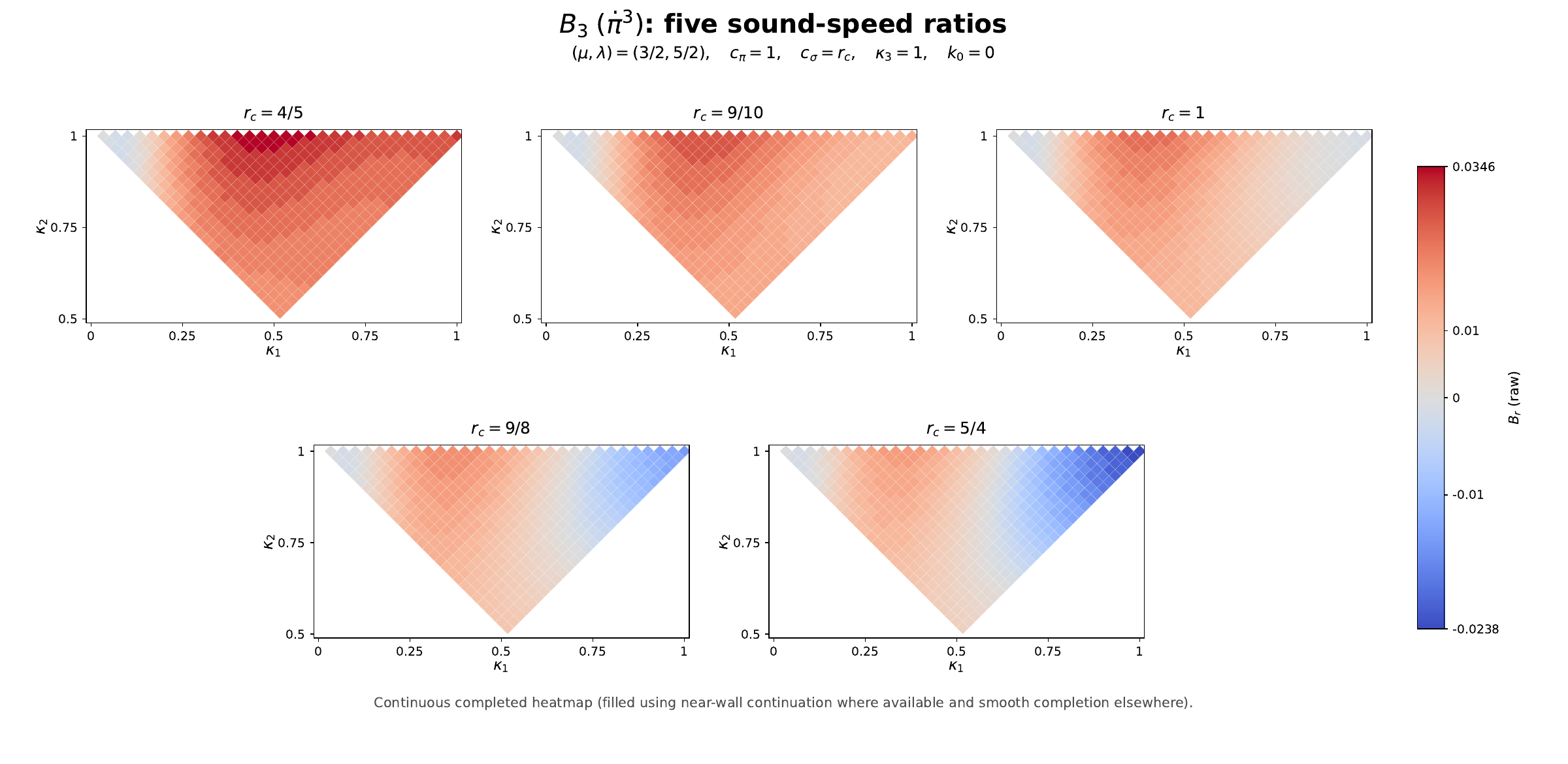}
\caption{Bispectrum $B_3$ arising from the $\dot{\pi}^3$
interaction for five sound-speed ratios
$r_c=\{4/5,\,9/10,\,1,\,9/8,\,5/4\}$. The remaining model parameters, momentum coordinates and panel ordering are as in figure~\ref{fig:bispectrum_rc45}.}
\label{fig:bispectrum_rc98}
\end{figure}

Figures~\ref{fig:bispectrum_rc45}--\ref{fig:bispectrum_rc98} show the momentum dependence of the four interaction channels. The three channels involving the entropy field exhibit a common dependence on the relative sound speed, whereas the purely adiabatic velocity interaction develops a distinct momentum and sign structure. The colour scales are chosen separately for each panel to resolve the variation within each channel.

For the three channels containing entropy-field insertions, the sampled coefficients have a definite sign: $B_0$ and $B_1$ are negative, while $B_2$ is positive. Their magnitudes decrease monotonically across the five sampled values of $r_c$ at every one of the $445$ momentum configurations shared by all five models at identical coordinates. For each sound speed, the largest sampled magnitudes of these three coefficients occur at the equilateral configuration.

The purely adiabatic velocity interaction behaves differently. Its coefficient $B_3$ takes both signs within the sampled momentum domain, and $166$ of the $445$ shared configurations attain opposite signs at different sound speeds. At the equilateral configuration, the sign changes between $r_c=9/10$ and $r_c=1$. The maximum positive value on the sampled grid lies along the isosceles edge $\kappa_2=\kappa_3=1$, moving from $\kappa_1=7/15$ at $r_c=4/5$ to $\kappa_1=1/3$ at $r_c=5/4$. The sound-speed dependence of this channel therefore cannot be absorbed into an overall rescaling of the bispectrum.

Table~\ref{tab:equilateral_bispectrum_rc} collects the equilateral values. Of the three channels containing entropy-field insertions, the mixed interaction $(\dot\pi/H)\sigma^2$ exhibits the strongest suppression, with its magnitude decreasing by a factor of approximately $7.3$ between $r_c=4/5$ and $r_c=5/4$. These changes in amplitude and momentum dependence occur at fixed mass and mixing parameters, and hence at fixed late-time scaling exponents.

\begin{table}[htbp]
\centering
\small
\begin{tabular}{crrrr}
\toprule
$r_c$ & $B_0$ & $B_1$ & $B_2$ & $B_3$ \\
\midrule
$4/5$  & $-0.332348$ & $-0.259588$ & $0.528878$ & $0.028979$ \\
$9/10$ & $-0.224719$ & $-0.165971$ & $0.440563$ & $0.012207$ \\
$1$    & $-0.158104$ & $-0.107475$ & $0.372061$ & $-0.001007$ \\
$9/8$  & $-0.106534$ & $-0.062483$ & $0.305872$ & $-0.013885$ \\
$5/4$  & $-0.074718$ & $-0.035448$ & $0.254927$ & $-0.023786$ \\
\bottomrule
\end{tabular}
\caption{Equilateral values of the reduced bispectrum coefficients \eqref{eq:bispectrumdefinition}, ordered by the number $r$ of velocity insertions in \eqref{eq:S3}, for $(\mu,\lambda)=(3/2,5/2)$ and $c_\pi=1$.}
\label{tab:equilateral_bispectrum_rc}
\end{table}

\subsection{Numerical accuracy}
\label{sec:errorbudget}

The auxiliary transport uses Taylor expansion orders $48$ and $60$ for the primary and reference solutions, with a target relative tolerance of $10^{-20}$. To check the construction of the differential system, we also evaluate the connection by direct matrix inversion rather than spectral decomposition at three representative momentum configurations with $\ep=1/10$. Starting from the same Gamma-series boundary prescription, the two implementations give integral vectors and assembled vertex values that agree to within a relative difference of $6\times10^{-56}$.

At each momentum configuration, we perform the calculation for twelve regulator values,
\begin{align}
\ep^{-1}\in\{10,12,16,20,24,32,40,48,64,80,96,128\}.
\label{eq:regulatorsamples}
\end{align}
The physical coefficient $B_r$ is obtained by fitting $2\,\operatorname{Im}I_r(\ep;\bm\kappa)$ with a fourth-degree polynomial and evaluating it at $\ep=0$. To assess the stability of this extrapolation, we compare its value $b_4$ with the results of a cubic fit, $b_3$, and a fourth-degree fit restricted to $\ep\leq1/24$, denoted by $b_{4,\mathrm{low}}$. We define the relative spread
\begin{align}
\delta_{\rm fit}=\frac{\max\left\{|b_3-b_4|,\ |b_{4,\mathrm{low}}-b_4|\right\}}{|b_4|}.
\label{eq:fitspread}
\end{align}

Across the $9220$ bispectrum values, the median relative spread is $1.32\times10^{-7}$ and the maximum is $7.57\times10^{-4}$. The $88$ values with relative spreads exceeding $10^{-5}$ all belong to the purely adiabatic velocity channel, where the small amplitudes near its zeros enhance the relative sensitivity of the extrapolation. The largest absolute departure from the reference fit across all four channels is $8.3\times10^{-8}$.

The boundary evaluation, differential-equation transport and algebraic reduction use complex ball arithmetic, with midpoint values passed between these stages. The regulator fits are performed in double precision. The transport and fit comparisons provide numerical stability diagnostics, while the accuracy of the physical integrals is tested independently by direct time integration in section~\ref{sec:validation}.

\section{Verification}
\label{sec:validation}

We test the numerical results by direct time integration and use exact representations from the literature to check the normalization of the coupled modes and the assembly of the correlators.

\subsection{Independent numerical checks}
\label{sec:directverification}

For the direct calculation, we initialize the Bunch--Davies subspace using a large-$z$ asymptotic expansion and integrate the original coupled equations \eqref{eq:coupled}, imposing the canonical normalization on the real axis. The response is evaluated on the Euclidean contour by propagating its decaying subspace with successive QR orthogonalizations. We integrate the cubic vertices by Gaussian quadrature, evaluating the contribution near the origin through the Frobenius expansion. This provides a reference calculation independent of the auxiliary-energy transport and IBP reduction.

We first compare the regulated vertex integrals at $c_\pi=c_\sigma=1$, $(\mu,\lambda)=(3,1)$ and $\bm\kappa=(1,1,1/2)$. At $\ep=1/10$ and $1/20$, the four complex integrals $I_r(\ep;\bm\kappa)$ obtained by transport to $k_0=0$ agree with direct contour integration to within a relative difference of $1.9\times10^{-7}$.

For the bispectrum surfaces of section~\ref{sec:surface}, we select $86$ momentum configurations across the five sound speeds, including equilateral triangles, points near the squeezed and folded boundaries, configurations close to frequency hypersurfaces, and points with relatively large extrapolation spreads. Increasing the quadrature order from $48$ to $64$ or varying the cutoff and Frobenius truncation changes the resulting coefficients $B_r$ by at most $2.4\times10^{-13}$. We also repeat the calculation at $17$ selected configurations with tighter mode-function integration tolerances, higher-order asymptotic initial data and increased quadrature orders. Table~\ref{tab:directverification} compares the resulting $344$ bispectrum coefficients with those obtained from the IBP--DE construction.

\begin{table}[htbp]
\centering
\small
\begin{tabular}{lcc}
\toprule
Interaction & Maximum absolute difference & Maximum relative difference \\
\midrule
$\sigma^3$ & $8.61\times10^{-11}$ & $1.75\times10^{-9}$ \\
$(\dot\pi/H)\sigma^2$ & $3.10\times10^{-10}$ & $6.52\times10^{-8}$ \\
$(\dot\pi/H)^2\sigma$ & $7.06\times10^{-9}$ & $3.51\times10^{-7}$ \\
$(\dot\pi/H)^3$ & $9.30\times10^{-8}$ & $4.27\times10^{-5}$ \\
\bottomrule
\end{tabular}
\caption{Comparison of the reduced bispectrum coefficients $B_r$ defined in \eqref{eq:bispectrumdefinition} with independent direct time integration at $86$ momentum configurations across the five sound speeds. Relative differences are measured with respect to the IBP--DE results.}
\label{tab:directverification}
\end{table}

The two calculations agree in sign at all sampled points for each of the four channels. The largest differences occur in the purely adiabatic velocity contribution, consistent with its greater sensitivity to regulator extrapolation. In particular, the largest relative difference occurs near the squeezed boundary, where the magnitude of the coefficient is of order $10^{-4}$.

\subsection{Comparison with exact representations}
\label{sec:literatureverification}

The power spectrum tests the Bunch--Davies normalization entering the Gamma-series boundary. For $c_\pi=1$, the enhancement relative to the unmixed single-field power spectrum is $2P$, with $P$ defined in \eqref{eq:response}. We extract $P$ from the coefficient of $z^{-1}$ in the production series for $W_1$ and compare it with the Heun-function expression of \cite{Huenupi:2026jcv}, using its Gamma-function limit at equal sound speeds. Across the five models of \eqref{eq:soundspeedscan}, the relative differences are below $1.8\times10^{-11}$.

We further test the direct integration procedure against two exact bispectrum representations at equal sound speeds. The squeezed $\sigma^3$ bispectrum of \cite{Wang:2026lff} involves two hard integrals $J^\pm_{\lambda,\rho}$, with their clock parameter $\rho$ corresponding to our $\gamma$. Accounting for the conjugate Wightman ordering of their transfer function, these integrals are related to our two-leg family by
\begin{align}
J^\pm_{\lambda,\gamma}
=\overline{\mathcal I^{(0)}_{(2,2)}\left(\tfrac12\mp i\gamma;(1,1)\right)}.
\label{eq:wangcomparison}
\end{align}
For $c_\pi=c_\sigma=1$ and $(\lambda,\gamma)=(1/2,1)$, direct integration gives
\begin{align}
\left|J^+-\overline{J^-}\right|&=0.04165165538,\qquad
\arg\left(J^+-\overline{J^-}\right)=1.336584733 .
\end{align}
The complex value agrees with the numerical evaluation of the double-integral representation in appendix D of \cite{Wang:2026lff} to a relative difference of $4.6\times10^{-10}$.

For the velocity interaction, we compare with the exact bispectrum representation of \cite{Pinol:2026xnl}. With the normalization of section~\ref{sec:numericalsetup}, the equilateral-normalized shape along the isosceles line is
\begin{align}
R(q)=\frac{B_3(q,1,1)}{qB_3(1,1,1)} .
\label{eq:pinolcomparison}
\end{align}
At $c_\pi=c_\sigma=1$ and $(\lambda,\gamma)=(1/2,5/2)$, we obtain $R(1/2)=0.8699491135$ and $R(1/10)=0.2997920729$. Over twelve ratios in the range $0.05\leq q\leq0.5$, the absolute differences between direct integration and the numerical evaluation of Eq.~(13) of \cite{Pinol:2026xnl} are below $6\times10^{-14}$.

\section{Summary and outlook}
\label{sec:summary}

We have extended the generalized IBP method for de Sitter correlators to strongly mixed propagation. A rational transformation of the coupled mode functions yields a four-component basis whose differential equation contains only a constant matrix and a simple pole. This structure leads to an iterative reduction of single-vertex integrals with an arbitrary number of legs and to closed differential equations in $\nd\log$ form. The frequency eigenvalues of the one-leg system determine the alphabet, and the momentum-space connection at each fixed weight decomposes into two independent sectors of equal dimension.

A central achievement of our approach is its systematic
and universal character. For an arbitrary single-vertex
integral family with $N$ legs, we have derived a universal
recursive reduction formula, \eqref{eq:Apm}, which
expresses integrals of different time weights in terms
of a finite basis of $4^N$ master integrals. These master
integrals satisfy a closed system of differential equations
in $\nd\log$ form, \eqref{eq:dlog}, whose structure
is determined entirely by the single-leg matrices and
their associated frequency projectors. Crucially, we
have also established a systematic prescription for
determining the boundary conditions from the
Bunch--Davies mode functions, using either the auxiliary
exponential construction or a direct momentum-space
anchor. The combination of the universal IBP reduction,
the general $\nd\log$ differential equations, and the
boundary-condition prescription therefore provides
a complete computational framework for evaluating the
single-vertex master-integral family. Since arbitrary
local interactions expressible in this family can be
assembled from the same master integrals, the method
is not restricted to the particular cubic vertices
investigated here, but offers a general and reusable
approach to strongly mixed cosmological correlators.

The boundary data follow from the canonically normalized Bunch--Davies response. Introducing an auxiliary energy turns its Frobenius expansion into a convergent Gamma-function series, from which the physical integrals are obtained by differential-equation transport and IBP reduction. A direct construction based on Frobenius subtraction provides an alternative boundary at a momentum anchor. The spectral structure also determines the indicial exponents at generic points of the frequency hypersurfaces. Unequal sound speeds introduce additional singularities of the differential connection inside the momentum triangle, although the Bunch--Davies contact integrals remain regular there.

We have applied this construction to the four cubic interactions of a strongly mixed two-field model, obtaining bispectrums over momentum configurations across five relative sound speeds. Across the five sampled sound speeds, the magnitudes of the three contributions involving the entropy field decrease with increasing $r_c$ throughout the common momentum grid. The purely adiabatic velocity interaction instead exhibits changes in both its sign structure and momentum dependence. The numerical results agree with independent time integration at representative configurations throughout the sampled domain.

A natural extension is to diagrams with several interaction vertices, where the time-ordering functions of the dressed bulk-to-bulk propagators generate additional terms in the IBP relations. The explicit alphabet and residue matrices may also provide a starting point for constructing multivariate analytic solutions, extending the hypergeometric representation of unmixed de Sitter correlators to strongly mixed theories.


\acknowledgments
Jiaqi Chen is supported by the National Natural Science Foundation of China (NSFC) under
Grant No.~12505094 and by the Science Foundation of China University of Petroleum, Beijing
(No.~2462025YJRC019).

The authors developed the scientific ideas and methodology. Kimi-3.1 Pro, GLM-5.3, Qwen-3.8 Max, Qwen-4 Max, DeepSeek-4.1 Flash, GPT-5.6 Sol, GPT-6 Sol and GPT-6 Astra (in no particular order) in codex and qoder assisted
with part of the code implementation and preparation of the paper draft.
The authors are responsible for the scientific content,
calculations, and conclusions.

\appendix
\section{Auxiliary-energy transport and weight reduction}
\label{app:k0}

The auxiliary construction of section~\ref{sec:auxexp} involves two operations: transport in the auxiliary energy and reduction in the time weight. We collect here the singularity structure of the auxiliary connection and the conditions under which the weight reduction is performed.

\subsection{Singularities of the auxiliary connection}

At fixed momenta, the spectral inverse \eqref{eq:inverseoperators} gives
\begin{align}
\det(A_\Sigma+ik_0\mathbb I_{4^N})
=i^{4^N}\prod_{\bm\omega}\left(k_0+\sum_{j=1}^N\omega_j\kappa_j\right),
\label{eq:k0det}
\end{align}
where $\omega_j$ are defined in \eqref{eq:frequencyspectrum} and the product runs over their tagged combinations. Writing \eqref{eq:auxiliaryconnection} as $\partial_{k_0}\bm J^{(n)}=\Omega_{k_0}(w)\bm J^{(n)}$, the auxiliary connection takes the $\nd\log$ form
\begin{align}
\Omega_{k_0}(w)\,\nd k_0
=-\sum_{\bm\omega}P_{\bm\omega}M_w\,
\nd\log\left(k_0+\sum_{j=1}^N\omega_j\kappa_j\right),
\label{eq:k0dlog}
\end{align}
with possible simple poles at
\begin{align}
k_0=-\sum_{i=1}^N\omega_i\kappa_i .
\label{eq:k0poles}
\end{align}
At a pole $k_0=\xi$, the residue is
\begin{align}
\mathcal R_\xi=-\mathcal P_\xi M_w,\qquad
\mathcal P_\xi=\sum_{\bm\omega:\,\sum_j\omega_j\kappa_j=-\xi}P_{\bm\omega}.
\label{eq:k0residue}
\end{align}
Here the projectors are grouped according to coincident frequency sums at the chosen momenta, rather than proportional frequency letters as in the momentum-space connection.

The residue spectra follow from the compression argument of section~\ref{sec:nspectrum}. Let $d_\xi=\operatorname{rank}\mathcal P_\xi$. At unequal sound speeds, distinct frequency states with the same frequency sum differ in at least two tensor slots, since all $\kappa_i$ are strictly positive. The compression of $B_\Sigma$ to the image of $\mathcal P_\xi$ therefore vanishes, giving $d_\xi$ exponents equal to $-w$ and $4^N-d_\xi$ zero exponents. At equal sound speeds, each sign pattern contributing to $\mathcal P_\xi$ carries the tensor sum of the blocks \eqref{eq:degeneratefrequencyblock}. If $s_\xi$ sign patterns contribute, the exponents on this subspace are
\begin{align}
\nu_j=-w+\frac{i\lambda}{2}(N-2j),\qquad
j=0,\ldots,N,
\label{eq:k0spectrum}
\end{align}
with multiplicities $s_\xi\binom Nj$, where $d_\xi=2^Ns_\xi$. The remaining $4^N-d_\xi$ exponents are zero.

For real positive momenta, the poles of the auxiliary connection lie on the real axis, and transport through the upper half plane follows the analytic branch of \eqref{eq:Jfamily}. The endpoint $k_0=0$ is an ordinary point whenever all frequency letters are nonzero. Even when the connection is singular there, the Bunch--Davies integral is analytic in $k_0$ near the origin, as established in section~\ref{sec:auxexp}. This property is preserved under meromorphic continuation in the weight away from its poles. Projecting the IBP relation \eqref{eq:JIBP} at $k_0=0$ onto the zero-frequency subspace gives the regularity condition
\begin{align}
\mathcal P_0M_w\bm J^{(n)}(0)=0,
\label{eq:k0regularity}
\end{align}
at weights where the meromorphically continued integral is regular.

\subsection{Descent to the physical weights}

We construct the Gamma-series boundary at an integer reference weight $n_*$ satisfying ${\rm Re}(\ep+n_*)>N$. After transport to the desired auxiliary energy, the lower weights follow from \eqref{eq:JIBP}:
\begin{align}
\bm J^{(n-1)}
=-M_{\ep+n-1}^{-1}(A_\Sigma+ik_0\mathbb I_{4^N})\bm J^{(n)} .
\label{eq:descent}
\end{align}
At $k_0=0$, this reduces to the downward shift of the physical integral family.

The spectrum \eqref{eq:weightspectrum} determines the singular weights of this recursion. Since $\gamma>0$, a real eigenvalue of $B_\Sigma$ must contain equal numbers of the conjugate exponents $-\tfrac12\pm i\gamma$. Each conjugate pair contributes $-1$, so all real eigenvalues of $B_\Sigma$ are integers. Taking a real regulator $0<\ep<1$ therefore keeps the weight matrices invertible throughout the integer-weight descent. At three legs, the steps producing the sectors $n=0,2,3$ become singular at $\ep=0$.

For the numerical results of section~\ref{sec:results}, we take $n_*=4$, transport to $k_0=0$ and descend successively to $n=3,2,1,0$. The full integral vectors can have poles in $\ep$, but the components entering the cubic vertices are regular at $\ep=0$. Indeed, a term containing $m$ factors of $W_3$ occurs at weight $n=m$ in \eqref{eq:velocity}, and its integrand satisfies
\begin{align}
t^{\ep+m-1}\prod_{i\in M}W_3(\kappa_i t)\prod_{j\notin M}W_2(\kappa_j t)
=O\left(|t|^{{\rm Re}\,\ep+1/2}\right),\qquad |M|=m ,
\label{eq:vertexconvergence}
\end{align}
near the origin along the integration contour. Each such term is integrable in a neighbourhood of $\ep=0$. In the numerical implementation, the regulator is retained during the descent, and the assembled vertex coefficients are extrapolated to zero as described in section~\ref{sec:errorbudget}.

\section{Laurent coefficients in the anchor construction}
\label{app:ladder}

The direct boundary construction of section~\ref{sec:anchor} allows us to work with the Laurent coefficients of the integral family rather than evaluating it at finite regulator values. The IBP relations and momentum-space differential equations close on the residues and finite parts.

\subsection{Upward weight recursion}

Since the one-leg Frobenius expansions contain no logarithms, the subtraction formula \eqref{eq:anchorfp} gives a meromorphic continuation in the weight with at most simple poles. Near an integer weight, we write
\begin{align}
\bm I^{(n)}(\ep;\bm\kappa)
=\frac{\bm R_n(\bm\kappa)}{\ep}+\bm F_n(\bm\kappa)+O(\ep),
\label{eq:laurentsector}
\end{align}
where $\bm R_n$ and $\bm F_n$ are the residue and finite part. If the product of the one-leg functions has the expansion $\sum_p\bm C_p t^p$, with equal powers collected, then $\bm R_n=\bm C_{-n}$, and the corresponding subtraction term contributes $\bm C_{-n}\log t_c$ to the finite part.

The possible poles of the integral vector depend on the full Frobenius series, including its subleading powers. They need not coincide with the singular weights of the downward IBP relation. For instance, the product of three $W_1$ components generically contains a $t^{-1}$ term, giving a pole at $w=1$ even though $M_1$ is invertible at three legs.

Away from the frequency hypersurfaces, the upward IBP relation can be written as
\begin{align}
\bm I^{(n+1)}
=-A_\Sigma^{-1}(M_n+\ep\mathbb I_{4^N})\bm I^{(n)},
\qquad M_n=n\mathbb I_{4^N}+B_\Sigma .
\label{eq:upladder}
\end{align}
Inserting \eqref{eq:laurentsector} and collecting the coefficients of $\ep^{-1}$ and $\ep^0$ gives
\begin{align}
\bm R_{n+1}&=-A_\Sigma^{-1}M_n\bm R_n,\nonumber\\
\bm F_{n+1}&=-A_\Sigma^{-1}(M_n\bm F_n+\bm R_n).
\label{eq:ladderFR}
\end{align}
The explicit regulator dependence of the shift matrix couples the residue of one sector to the finite part of the next. Starting from $\bm R_0$ and $\bm F_0$ at an ordinary momentum configuration, these relations determine the Laurent coefficients at all higher integer weights without inverting the weight matrix.

\subsection{Momentum transport}

The momentum-space connection has an analogous structure. Since \eqref{eq:dematrix} is affine in the weight,
\begin{align}
\Omega_{\kappa_i}(n+\ep)
=\Omega_{\kappa_i}(n)-\ep A^{(i)}A_\Sigma^{-1},
\end{align}
the Laurent coefficients satisfy
\begin{align}
\partial_{\kappa_i}\bm R_n
&=\Omega_{\kappa_i}(n)\bm R_n,\nonumber\\
\partial_{\kappa_i}\bm F_n
&=\Omega_{\kappa_i}(n)\bm F_n-A^{(i)}A_\Sigma^{-1}\bm R_n .
\label{eq:laurenttransport}
\end{align}
Using the spectral representation \eqref{eq:dlog}, we obtain the corresponding $\nd\log$ system,
\begin{align}
\nd\bm R_n&=\Omega(n)\bm R_n,\nonumber\\
\nd\bm F_n&=\Omega(n)\bm F_n
-\sum_{\bm\omega}P_{\bm\omega}\bm R_n\,
\nd\log\left(\sum_{j=1}^N\omega_j\kappa_j\right).
\label{eq:laurentdlog}
\end{align}
The residue and finite part thus form a closed triangular differential system with the same alphabet as the original integral family.

The reference coefficients $\bm R_0$ and $\bm F_0$ can be transported from the anchor using \eqref{eq:laurentdlog}, after which the higher weights follow from \eqref{eq:ladderFR}. The cubic contributions are assembled from the finite parts using \eqref{eq:velocity} and \eqref{eq:bispectrumdefinition}. This gives an alternative to performing the transport and reduction at finite $\ep$.


\bibliographystyle{JHEP}
\bibliography{references2}

\end{document}